\documentclass[aps,prb,twocolumn,showpacs]{revtex4}

\usepackage{graphicx}
\usepackage{amsmath} \newcommand{\EqLabel}[1]{\label{#1}}
\newcommand{\mb}[1]{\mathbf{#1}} \newcommand{\mr}[1]{\mathrm{#1}}

\begin{document}
 
 \title{Momentum average approximation for models
 with electron-phonon coupling dependent on the phonon momentum}

\author{Glen L. Goodvin and Mona Berciu}

\affiliation{Department of Physics and Astronomy, University of
  British Columbia, Vancouver, British Columbia, Canada V6T 1Z1}

\date{\today}
 
\begin{abstract}
We generalize the momentum average (MA) approximation to study the
properties of models with $g_{\mathbf{q}}$ momentum-dependent electron-phonon
coupling. As in the case of the application of the original MA to the Holstein
model, the results are analytical, numerically trivial to evaluate,
exact for both zero bandwidth and for zero electron-phonon coupling,
and are accurate everywhere in parameter space. Comparison with
available numerical data confirms this accuracy. We then show that
further improvements can be obtained based on variational
considerations, using the one-dimensional breathing-mode Hamiltonian
as a specific example. For example, by using this variational MA, we
obtain ground state energies within at most 0.3\% error of the numerical data.
\end{abstract}

 \maketitle

\section{Introduction}
There is considerable need to understand the coupling of a
particle to its environment, in particular the interaction between a
charge carrier and phonons. The physics of polarons -- electrons
dressed by phonon clouds -- is believed to be relevant in
explaining a plethora of physical systems and properties, for
example in polymers, nanotubes, C$_{60}$ and other
fullerenes,\cite{hengsberger,gunnarsson,su} manganites,\cite{salamon}
Bechgaard salts,\cite{mila,magna} and even, possibly, 
the
 kink in the quasiparticle dispersion in cuprates, a necessary
 ingredient for the 
full understanding of high-temperature
superconductivity.\cite{lanzara,shen,mishchenko,cuk} 

It is of obvious benefit to have highly accurate analytical
approximations that can be applied to complicated many-body problems,
such as the polaron problem. Unfortunately, most existing approximations
fail to reproduce the correct physics in ``intermediate'' regimes
where the relevant physics often occurs. In recent
works\cite{berciu1,goodvin, berciu3, berciu2, covaci, covaci2} we have
proposed the 
so-called momentum average (MA) approximation for calculating the
Green's function of a dressed particle, focusing on the case of
an electron dressed by phonons. This approximation is analytical,
easy-to-use, and highly accurate throughout all of parameter
space, both at low and at high-energies. Its underlying idea  is to sum
all of the 
diagrams in the diagrammatical expansion of the self-energy, but with
the diagrams approximated in such a way that the full summation can be
performed. At the simplest level, this is achieved by replacing all of
the free propagators 
appearing in the self-energy by their momentum average. For the
Holstein model, the resulting MA self-energy exhibits not only the
exact asymptotic behaviour for zero coupling and zero bandwidth, but
it is also in excellent agreement with numerical results in the
intermediate regimes where other approximate methods completely
fail.\cite{berciu1,goodvin,berciu2}

In Refs. \onlinecite{berciu1} and \onlinecite{goodvin} we presented a
detailed account of the derivation and application of the
simplest-level MA
approximation (now known as MA$^{(0)}$)  to the Holstein model,
including extensive comparisons to 
the available numerical data. We further justified the accuracy of the
approximation by looking at the spectral weight sum rules. The MA$^{(0)}$
approximation satisfies the first six spectral weight sum rules
exactly, and  is highly accurate for all higher order sum rules. In 
Ref. \onlinecite{goodvin} we also pointed out the three key limitations
of the MA$^{(0)}$ approximation: (i) it fails to correctly predict the
so-called 
electron+phonon continuum that must occur at a phonon energy above the
ground state energy of the polaron, (ii) its accuracy worsens for
small phonon energies, and (iii) the MA$^{(0)}$ self-energy is independent of
momentum. 

These shortcomings motivated the systematic improvement of
the MA approximation, presented in Ref. \onlinecite{berciu2}. This
lead to a hierarchy of increasingly more accurate approximations,
which we call MA$^{(0)}$, MA$^{(1)}$, MA$^{(2)}$, etc. The increased
accuracy is a result of fewer approximations for the self-energy
diagrams, however done in such a way that we can still
sum all the resulting diagrams analytically. It is also possible to
understand the  MA approximations in a variational context, as detailed
below. From this point of view, the systematic  improvements are
obtained by including 
additional states required to reproduce correctly the electron+phonon
continuum. In the process we also obtain a momentum-dependent self-energy
and higher accuracy throughout all of parameter space, particularly
for small phonon energies. As a result, the MA approximations allow us
to understand very accurately the Holstein polaron physics, throughout
the parameter space, for all energies and momenta. 

In this paper we generalize this powerful set of 
approximations to a much broader class of
models with electron-phonon coupling that depends on the phonon
momentum. We first derive a simple generalization, leading to what we
will continue to call the  MA$^{(0)}$, MA$^{(1)}$, MA$^{(2)}$,
etc. hierarchy. These are very easy to apply to any Hamiltonian of
this class, however while still asymptotically exact for both weak and
strong coupling, at intermediary couplings the relative errors are of a
few percent. In other words, these can be used to get a quick estimate
of typical energies and spectra. We then show how these can be
significantly improved, using variational considerations. This generates a
second hierarchy of approximations which we call  MA$^{(v,0)}$,
MA$^{(v,1)}$, MA$^{(v,2)}$, 
etc., with relative errors well below one percent. As a test case to
gauge these accuracies, we use  the one-dimensional (1D) 
breathing-mode Hamiltonian, where high-accuracy numerical
results have  recently  become available.\cite{lau} 

The general electron-phonon coupling model that we consider has the
following form in momentum space:
\begin{equation} \label{eq:genH}
 {\cal H} = \sum_{\mb{k}} \varepsilon_{\mb{k}} c_{\mb{k}}^{\dagger}
c_{\mb{k}} + \Omega \sum_{\mb{q}} b_{\mb{q}}^{\dagger}b_{\mb{q}} +
\sum_{\mb{k},\mb{q}} \frac{g_{\mb{q}}}{\sqrt{N}}
c_{\mb{k}-\mb{q}}^{\dagger} c_{\mb{k}} \left( b_{\mb{q}}^{\dagger} +
b_{-\mb{q}} \right).
\end{equation}
The first term is the kinetic energy of the electron, where
$c_{\mb{k}}^{\dagger}$ and $c_{\mb{k}}$ are electron creation
and annihilation operators, and $\varepsilon_{\mb{k}}$ is the electron
dispersion. For the single electron (polaron) problem of interest to
us, the spin of the electron is irrelevant and we suppress its
index. The second term describes a branch of optical phonons of energy
$\Omega$, $b_{\mb{q}}^{\dagger}$ and $b_{\mb{q}}$ being the
phonon creation and annihilation operators.  The last term describes
the coupling of the electron to the phonons, where $g_{\mb{q}}$ is the
momentum-dependent coupling. Sums are over all momenta inside the
first Brillouin zone, and we set
$\hbar=1$ and $a=1$ throughout this paper.
This general Hamiltonian covers complicated
electron-phonon couplings such as those found in the 
Rashba-Pekkar,\cite{rashba,dykman} Fr\"ohlich,\cite{frohlich} and
breathing-mode\cite{slezak,lau} models, and it reduces to the Holstein
model when $g_{\mb q}$ is simply a constant.\cite{holstein} 

The quantity of interest to us is the (retarded) single polaron
Green's function:\cite{berciu1, goodvin}
\begin{equation} \label{eq:Gdef}
G(\mb{k}, \omega) = \langle 0 | c_{\mb{k}} \hat{G}(\omega)
c_{\mb{k}}^\dag|0\rangle = \langle 0 | c_{\mb{k}} {1\over \omega -
{\cal H} + i \eta}c_{\mb{k}}^\dag|0\rangle,
\end{equation}
where $|0\rangle$ is the vacuum, $c_{\mb{k}}|0\rangle = b_{\mb{q}}
|0\rangle =0$, and $\eta > 0 $ is infinitesimally small. The 
importance of this Green's function is obvious in the Lehmann
representation:\cite{mahan}
\begin{equation}
\EqLabel{leh} G(\mb{k},\omega) = \sum_{\alpha}^{} \frac{|\langle
\alpha | c_{\mb{k}}^\dag|0\rangle|^2}{\omega - E_\alpha + i\eta},
\end{equation}
where $\{ |\alpha\rangle\}$ and $\{E_\alpha\}$ are the complete set of
one-particle eigenstates and eigenenergies, ${\cal H}|\alpha\rangle
= E_\alpha |\alpha\rangle$. In this representation it is clear that
the poles of the Green's function give the  one-particle
spectrum, and the associated residues, sometimes called quasiparticle (qp)
weights, give partial information on the nature of the
eigenstates. Knowledge of this Green's function  also allows us to
calculate other relevant quantities such as the effective mass of the
polaron, the average number of
phonons in the polaronic cloud,\cite{berciu1, goodvin} or more detailed
phonon numbers statistics.\cite{berciu4} Furthermore, its
imaginary part is 
directly measured experimentally by angle-resolved photoemission
spectroscopy (ARPES).\cite{damascelli}

In this paper we will use the 1D breathing-mode Hamiltonian as an
example, however, we stress that the methods we present  are
applicable to any Hamiltonian like Eq. (\ref{eq:genH}) in any
dimension. The reason for choosing the 1D breathing-mode
Hamiltonian is two-fold. First, in its full 2D form, it
describes lattice vibrations in a CuO$_2$-like plane, where the motion
of the O ions living on the bonds connecting the Cu sites is the
most important vibrational degree of freedom,\cite{lau} making it
possibly relevant for the study of 
high-T$_c$ superconductors. 
%
%
The second reason is that exact
diagonalization (ED) results\cite{lau} have recently become available for its
1D 
analog, relevant for CuO chains. We focus here on
this 1D  breathing-mode model because these results 
serve as an excellent gauge of 
the accuracy of the generalized MA approximations. Their availability is
very fortunate because although there are many numerical results for
the Holstein model, it is only due to recent advancements in
computational power that more complicated
electron-phonon coupling models, such as the breathing-mode
Hamiltonian, can be investigated numerically. We also mention that
this model has been studied using 
the self-consistent Born approximation (SCBA),\cite{slezak} but SCBA
is known to be very poor for intermediate and large coupling
strengths.\cite{goodvin,lau}

In the 1D breathing-mode model, one considers a chain with two
interlaced sublattices, where 
the Cu sites which host the electron are indexed by integer labels,
and the O sites which host the phonons are indexed
by half-integer labels. The interaction term of the breathing-mode
Hamiltonian can be written in real space as
\begin{equation} \label{eq:gi}
g \sum_i c_i^{\dagger} c_i \left( x_{i+1/2} - x_{i-1/2} \right),
\end{equation}
where $x_{i\pm 1/2} = b_{i\pm 1/2}^{\dagger} + b_{i\pm 1/2}$
describe
the displacements of the O atoms neighboring the Copper atom at
site $i$ that hosts the electron, and $g$ is a constant describing the
strength of the 
electron-phonon coupling ($g$ absorbs the proportionality factors
between the true displacement $x_{i\pm 1/2}$ and the phonon operator
$b_{i\pm 1/2}^{\dagger} + b_{i\pm 1/2}$). Transforming into momentum
space, the 1D 
breathing-mode Hamiltonian takes the form of Eq. (\ref{eq:genH}) with
\begin{equation} \label{eq:gq}
g_q = -2ig \sin{qa\over 2}.
\end{equation}
In this model the electron motion is described by a tight-binding
model with the usual  $\varepsilon_k=-2t \cos(k
a)$, although our results can be applied for any dispersion.

This article is organized as follows. In Sec. II, we derive the exact
set of equations giving the Green's function of a polaron, review the MA
approximations  and derive the simple straightforward
generalizations MA$^{(0)}$, MA$^{(1)}$, etc. for models similar to
Eq. (\ref{eq:genH}). In Sec. III we show how 
to obtain the more accurate MA$^{(v,0)}$ and MA$^{(v,1)}$
approximations based on variational ideas. Here we use the 1D
breathing-mode Hamiltonian 
as an explicit
example. In Sec. IV we present our results and compare them to
the available numerical data where possible. Finally, Sec. V contains
our summary and conclusions.

\section{Calculating the Green's Function}
\subsection{Exact Solution}
\begin{figure}[b]
\includegraphics[width=0.90\columnwidth]{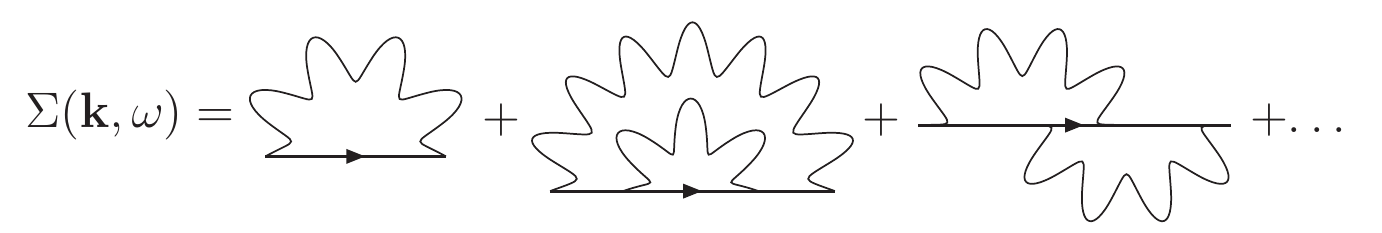}
\caption{Diagrammatical expansion of the
self-energy $\Sigma(\mb{k},\omega)$. \label{fig:expansion}}
\end{figure}

Although exact solutions for these types of problems are generally not
obtainable in 
closed form, one can  write down their formal solutions in terms
of an infinite set of coupled equations involving related
(higher-order) Green's functions.  In a previous
work\cite{goodvin} we described in detail how to generate these
equations for the Holstein 
model. The generalization to 
momentum-dependent coupling models is straightforward, by repeated use
of Dyson's identity $\hat{G}(\omega)=\hat{G}_0(\omega)+
\hat{G}(\omega)\hat{V}\hat{G}_0(\omega)$, where $\hat{G} = [\omega -
\hat{{\cal H}} + i\eta]^{-1}$, $\hat{G}_0=[\omega - \hat{{\cal H}}_0 +
i\eta]^{-1}$, $\hat{{\cal H}}=\hat{{\cal H}}_0+\hat{V}$, and $\hat{V}$
is the electron-phonon interaction. Following this procedure and
defining the
generalized Green's functions
\begin{equation}
\label{eq:FF}
F_n(\mb{k}, \mb{q}_1,\dots,\mb{q}_n, \omega)=\langle0| c_{\mb{k}}
\hat{G}(\omega)c_{\mb{k}-\mb{q}_T}^{\dagger}
b_{\mb{q}_1}^{\dagger}\dots b_{\mb{q}_n}^{\dagger}|0 \rangle
\end{equation}
where $F_0(\mb{k},\omega)=G(\mb{k},\omega)$, we find that
\begin{equation} \label{eq:G}
G(\mb{k},\omega) = G_0(\mb{k}, \omega) \left[
1+\frac{1}{\sqrt{N}}\sum_{\mb{q}_1} g_{\mb{q}_1}
F_1(\mb{k},\mb{q}_1,\omega) \right],
\end{equation}
and for $n \ge 1$,
\begin{widetext}
\begin{multline} \label{eq:F}
F_n(\mb{k}, \mb{q}_1, \ldots, \mb{q}_n, \omega) = \frac{1}{\sqrt{N}}
G_0(\mb{k}-\mb{q}_T,\omega-n\Omega) \left[ \sum_{i=1}^{n}
g_{-\mb{q}_i} F_{n-1} (\mb{k}, \ldots, \mb{q}_{i-1}, \mb{q}_{i+1},
\ldots, \omega) \right. \\ + \left. \sum_{\mb{q}_{n+1}}
g_{\mb{q}_{i+1}} F_{n+1}(\mb{k}, \mb{q}_1, \ldots, \mb{q}_{n+1},
\omega) \right].
\end{multline}
\end{widetext}
 The total momentum
carried by the phonons is denoted by ${\mb{q}_T} = \sum_{i=1}^{n}
\mb{q}_i$ and $G_0(\mb{k}, \omega) = (\omega - \epsilon_{\mb{k}} + i
\eta)^{-1}$ is the free electron Green's function. 
Observing from Eqs. (\ref{eq:F}) that all of these generalized Green's
functions 
$F_1,F_2,\ldots$ must be proportional to
$G(\mb{k},\omega)$,\cite{berciu2} we can recast our equations into a
more convenient form by defining
\begin{equation}
\label{eq:fF}
f_n(\mb{q}_1, \ldots, \mb{q}_n) = \frac{N^{n/2}g_{\mb{q}_1}\cdots
g_{\mb{q}_n} F_n(\mb{q}_1,\ldots,\mb{q}_n)}{G(\mb{k},\omega)},
\end{equation}
where we have also introduced the shorthand notation
$f_n(\mb{k},\mb{q}_1,\ldots,\mb{q}_n,\omega)\equiv
f_n(\mb{q}_1,\ldots,\mb{q}_n)$ (\emph{i.e.}, the $\mb{k}$ and $\omega$
dependence of these functions is implicitly assumed from now on).  In
this notation, Eq. (\ref{eq:G}) becomes 
\begin{equation} \label{eq:G'}
G(\mb{k},\omega) = G_0(\mb{k}, \omega) \left[
1+\frac{1}{N}\sum_{\mb{q}_1} f_1(\mb{q}_1) G(\mb{k},\omega) \right]
\end{equation}
with a solution written in the standard form
\begin{equation}
G(\mb{k},\omega) = \frac{1}{\omega - \varepsilon_{\mb{k}} -
\Sigma(\mb{k},\omega) + i\eta},
\end{equation}
where the self-energy is 
\begin{equation} \label{eq:Sigma}
\Sigma(\mb{k},\omega) = \frac{1}{N}\sum_{\mb{q}_1} f_1(\mb{q}_1).
\end{equation}
This self-energy can also be written in terms of an infinite set of
diagrams, as shown in Fig. \ref{fig:expansion}.

The quantity of interest $ f_1(\mb{q}_1)$ is obtained from the set of
 equations resulting from Eq. (\ref{eq:F}), namely $f_0 \equiv
1$, by definition, and for $n\ge 1$:
\begin{multline} \label{eq:f}
f_n(\mb{q}_1, \ldots, \mb{q}_n) = G_0(\mb{k}-\mb{q}_T,\omega-n\Omega)
\\ \times \left[ \sum_{i=1}^{n} |g_{\mb{q}_i}|^2 f_{n-1} (\ldots,
\mb{q}_{i-1}, \mb{q}_{i+1}, \ldots) \right. \\ +
\left. \frac{1}{N}\sum_{\mb{q}_{n+1}} f_{n+1}(\mb{q}_1, \ldots,
\mb{q}_{n+1}) \right].
\end{multline}

Of course this system can be solved trivially in the limit of $g=0$,
in which case $G(\mb{k},\omega)=G_0(\mb{k},\omega)$ directly from
Eq. (\ref{eq:G}). Also, in the limit of $t=0$ the free propagators
become independent of momentum and the equivalent of the Lang-Firsov result is
reproduced.\cite{goodvin} However, for the general case of finite $g$
and $t$, a closed form solution cannot be obtained, even for the
Holstein model.\cite{goodvin} Approximations are therefore needed.
We begin by describing the simplest MA$^{(0)}$ version of a
generalized MA approximation.

\subsection{The MA$^{(0)}$ Approximation \label{sec:MA0}}
For the Holstein
model, the MA$^{(0)}$ approximation amounts to replacing all of the
free electron propagators in the diagrammatical expansion of the
self-energy by their momentum average over the Brillouin zone:
\begin{equation} \label{eq:g0}
\bar{g}_0(\omega) = \frac{1}{N}\sum_q G_0(\mb{q},\omega),
\end{equation}
which is equivalent to replacing all $G_0(\mb{k} -
\mb{q}_T,\omega-n\Omega)$ by $\bar{g}_0(\omega-n\Omega)$ in
Eq. (\ref{eq:f}).  The procedure is essentially the same for
momentum-dependent electron-phonon couplings. We note that the first
term from the $n=1$ case of Eq. (\ref{eq:f}) does not actually require
any approximation (this is trivially true in the case in the Holstein
model) because $f_0 \equiv 1$ by definition, and its coefficient can
be written explicitly as
\begin{equation} \label{eq:g0barbar}
\bar{\bar{g}}_0(\mb{k},\omega) = \frac{1}{N}\sum_{\mb{q}}
|g_{\mb{q}}|^2G_0(\mb{k}-\mb{q},\omega).
\end{equation}
By making the substitution $G_0(\mb{q},\omega - n\Omega) \to
\bar{g}_0(\omega - n\Omega)$ everywhere else, we can rewrite
Eqs. (\ref{eq:Sigma}) and (\ref{eq:f}) in terms of the momentum
averaged functions ${\cal F}_n(\mb{k}, \omega) = {1\over N^n}
\sum_{\mb{q}_1,\dots,\mb{q}_n} f_n(\mb{q}_1, \ldots, \mb{q}_n)$. This
leads to simple recurrence relations linking each ${\cal F}_{n}$ to
${\cal F}_{n-1}$ and ${\cal F}_{n+1}$, and these can be solved in terms of
continued fractions.\cite{goodvin,berciu2} We find
\begin{equation} \label{eq:MA0}
\Sigma_{MA^{(0)}}(\mb{k},\omega) =
\frac{\bar{\bar{g}}_0(\mb{k},\omega-\Omega)}{1-\bar{g}^2\bar{g}_0(\omega
- \Omega)A_2(\omega)},
\end{equation}
where we have defined the infinite continued fractions
\begin{eqnarray}
\label{eq:An}
A_n (\omega) =&& \frac{n \bar{g}_0(\omega - n\Omega)}{1-\bar{g}^2
\bar{g}_0(\omega -n\Omega)A_{n+1}(\omega)} \\
\nonumber =&&\cfrac{n\bar{g}_0(\omega -
n\Omega)}{1-\cfrac{(n+1)\bar{g}^2\bar{g}_0(\omega -
n\Omega)\bar{g}_0(\omega - (n+1)\Omega)}{1- \cdots}},
\end{eqnarray}
and the momentum averaged electron-phonon coupling:
\begin{equation}
\bar{g}^2 = \frac{1}{N}\sum_{\mb{q}} |g_{\mb{q}}|^2.
\end{equation}
It is trivial to check that this reduces to the Holstein result of
Ref. \onlinecite{goodvin} when $g_{\mb{q}}$ is independent of
momentum.

Before we discuss how to systematically improve this result,
we briefly review one explanation as to why this is a reasonable first
step in obtaining an accurate approximation: in real space this
approximation is 
equivalent to replacing, in all self-energy diagrams,  all free
propagators $G_0(i,j,\omega - 
n\Omega)\rightarrow \delta_{i,j}\bar{g}_0(\omega - n\Omega)$, where $i$
and $j$ index the electron sites.\cite{barisic,berciu3} If the free
propagators $G_0(i,j,\omega-n\Omega)$ are evaluated for energies well
below the free electron continuum (and for a system with interactions
the polaron ground state is below the free
electron continuum), it is well known that the free propagator
decreases exponentially with increasing distance $|i-j|$. Thus,  the most
important terms are those with $i=j$, {\em i.e.} precisely those
included within 
MA$^{(0)}$, explaining why the approximation should be reasonably
accurate at low energies (spectral weight sum rules then insure that
it is similarly accurate for all energies). 

This also suggests a means to systematically improve the
MA$^{(0)}$ approximation. Since the propagators with the higher
energy (the energy closer to the free-electron continuum) decrease
exponentially with distance most slowly, the first improvement is to
keep all propagators with 
energy $\omega - \Omega$ exactly. Higher order systematic
improvements denoted by MA$^{(n)}$ would amount to keeping all
propagators with the argument $\omega - m\Omega$ exactly, for all $m \leq
n$. In Ref. \onlinecite{berciu2} we derived explicitly the
Holstein MA self-energy for both $n=1$ and $n=2$. As shown and
explained there, the
$n=1$ order already insures the key improvement of properly predicting
the polaron+one phonon continuum, which is usually absent at the $n=0$
level. The reason we went to $n=2$ for Holstein is that, given the
simplicity of that model, only at the MA$^{(2)}$ level did we find a
momentum-dependent self-energy. 

In contrast, for a model with a momentum-dependent coupling
even the MA$^{(0)}$ level gives explicit $\mb{k}$ dependence in
the self-energy [see Eq. (\ref{eq:MA0})], therefore in the next
subsection we only consider the MA$^{(1)}$ generalization.

\subsection{The MA$^{(1)}$ Approximation \label{sec:MA1}}
The main drawback of the MA$^{(0)}$ level of  approximation is
that it predicts an incorrect location for the electron+phonon
continuum that must start at $E_{GS}+\Omega$. This problem is always
``cured'' at the MA$^{(1)}$ level because in real space MA$^{(1)}$
includes states with a phonon cloud near the electron \emph{and} a
single phonon arbitrarily far away, in other words
precisely the type of 
states that give rise to the polaron+phonon continuum.\cite{berciu2}
Mathematically, MA$^{(1)}$ amounts to keeping all propagators with the
argument $\omega - \Omega$ in Eq. (\ref{eq:f}) exactly. Such terms
only appear in the $n=1$ equation:
\begin{equation} \label{eq:f1MA1}
f_1^{(1)}(\mb{q}_1) = G_0(\mb{k}-\mb{q}_1, \omega -
\Omega)\left[|g_{\mb{q}_1}|^2 + \frac{1}{N}\sum_{\mb{q}_2}
f_2^{(1)}(\mb{q}_1,\mb{q}_2) \right],
\end{equation}
where we have distinguished the approximated $f_n$ terms with the
superscript (1) to indicate the MA$^{(1)}$ level of
approximation.  For the remaining equations ($n\ge 2)$ we proceed
as before, 
replacing $G_0(\mb{k}-\mb{q}_T, \omega - n\Omega)$ with $g_0(\omega -
n\Omega)$ everywhere in Eq. (\ref{eq:f}):
\begin{multline} \label{eq:fnMA1}
f_n^{(1)}(\mb{q}_1,\ldots,\mb{q}_n) = \bar{g}_0(\omega - n\Omega) \\
\times \left[ \sum_{i=1}^n |g_{\mb{q}_1}|^2 f_{n-1}^{(1)}(\ldots,
\mb{q}_{i-1}, \mb{q}_{i+1}, \ldots) \right. \\ \left. + \frac{1}{N}
\sum_{\mb{q}_{n+1}}
f_{n+1}^{(1)}(\mb{q}_1,\ldots,\mb{q}_{n+1})\right].
\end{multline}
We wish to solve for $\Sigma_{MA^{(1)}}(\mb{k},\omega) = (1/N)
\sum_{\mb{q}_1} f_1^{(1)}(\mb{q}_1)$. The procedure is analogous to
that of  Ref. \onlinecite{berciu2}. We obtain two sets of
coupled recurrence relations, one for fully momentum averaged
quantities ${\cal F}_n(\mb{k}, \omega) = {1\over N^n}
\sum_{\mb{q}_1,\dots,\mb{q}_n} f^{(1)}_n(\mb{q}_1, \ldots, \mb{q}_n)$
that already appeared at MA$^{(0)}$ level, and one for partially
averaged quantities $\bar{\cal F}_n(\mb{q_1}, \mb{k}, \omega)
= {1\over N^{n-1}} 
\sum_{\mb{q}_2,\dots,\mb{q}_n} f^{(1)}_n(\mb{q}_1, \ldots,
\mb{q}_n)$. Their solution follows the procedure detailed in
Ref. \onlinecite{berciu2}, 
and we simply state the
result:
\begin{equation} \label{eq:MA1}
\Sigma_{MA^{(1)}}(\mb{k},\omega) =
\frac{\bar{\bar{g}}_0(\mb{k},\tilde{\omega})}
	 {1-\bar{\bar{g}}_0(\mb{k},\tilde{\omega})\left[   
A_2(\omega) - A_1(\omega - \Omega) \right]},
\end{equation}
where $\tilde{\omega} = \omega - \Omega - \bar{g}^2A_1(\omega -
\Omega)$. This expression is slightly more complicated than the
expression for $\Sigma_{MA^{(0)}}(\mb{k},\omega)$ because it involves
two continued fractions, but it is again trivial to compute
numerically.

\subsection{MA$^{(0)}$ and MA$^{(1)}$ Results \label{sec:MAresults}}
To illustrate the accuracy of MA$^{(0)}$ and MA$^{(1)}$ we will use
the 1D breathing-mode Hamiltonian described by Eqs. (\ref{eq:genH})
and (\ref{eq:gq}) as an example, comparing our results to those found
numerically using ED in  Ref. \onlinecite{lau}.

Before reporting these results we briefly discuss how the MA
results are evaluated in practice. First, the numerical evaluation of
the infinite continued fraction found in Eq. (\ref{eq:MA0}) requires
that we truncate the fraction at some level $n$. For an error of order
$\epsilon$ we require that at the $n^{\textrm{th}}$ level we have
$\epsilon > ng^2\bar{g}_0(\omega - n\Omega)\bar{g}_0(\omega -
(n+1)\Omega)$. Since $\bar{g}_0(\omega)\rightarrow 1/\omega$ when $|\omega| \to
\infty$, for large enough $n$ we can approximate $\bar{g}_0(\omega -
n\Omega) \approx \bar{g}_0(\omega - (n+1)\Omega) \approx
-1/(n\Omega)$. It then follows that for an error of order $\epsilon$,
we must have
\begin{equation}
n > \frac{1}{\epsilon}\frac{g^2}{\Omega^2}.
\end{equation}
In practice we always check our results by doubling the value of the
truncation value $n$ until the change in the self-energy is negligible. The
result is that all MA error bars throughout this paper are smaller
than the thickness of the lines/symbols used in the plots.

With our approximations for the Green's function in hand, the
ground-state properties are found by tracking the energy and weight of
the lowest pole of the spectral weight, defined as
\begin{equation}
A(\mb{k},\omega) = -\frac{1}{\pi} \textrm{Im}\,\, G(\mb{k},\omega) =
\sum_{\alpha} |\langle | \alpha | c_{\mb{k}}^{\dagger} | 0 \rangle |^2
\delta (\omega - E_{\alpha}),
\end{equation}
where a small but finite value of $\eta$ broadens the $\delta$ peaks
into Lorentzians and enables us to detect them numerically (typically
we take $\eta \sim 10^{-5}$). A  detailed description on how we
use the Green's function to extract the energy spectrum and
quasiparticle (qp) weights, as well as other quantities of interest
such as effective masses, average number of phonons in the polaron
cloud, etc., is presented in Ref. \onlinecite{goodvin}. The
explicit expressions for $\bar{g}_0(\omega)$ and
$\bar{\bar{g}}_0(k,\omega)$ needed  to compute the self-energies for
the 1D breathing-mode Hamiltonian are given in Appendix
\ref{sec:averages}. As is customary, we define the dimensionless
effective coupling $\lambda$ as the ratio between the lattice
deformation energy\cite{lau} $-2g^2/\Omega$ and the free electron
ground state energy $-2t$:
\begin{equation} \label{eq:lambda}
\lambda = \frac{g^2}{\Omega t}.
\end{equation}

\begin{figure}[t]
\includegraphics[width=0.98\columnwidth]{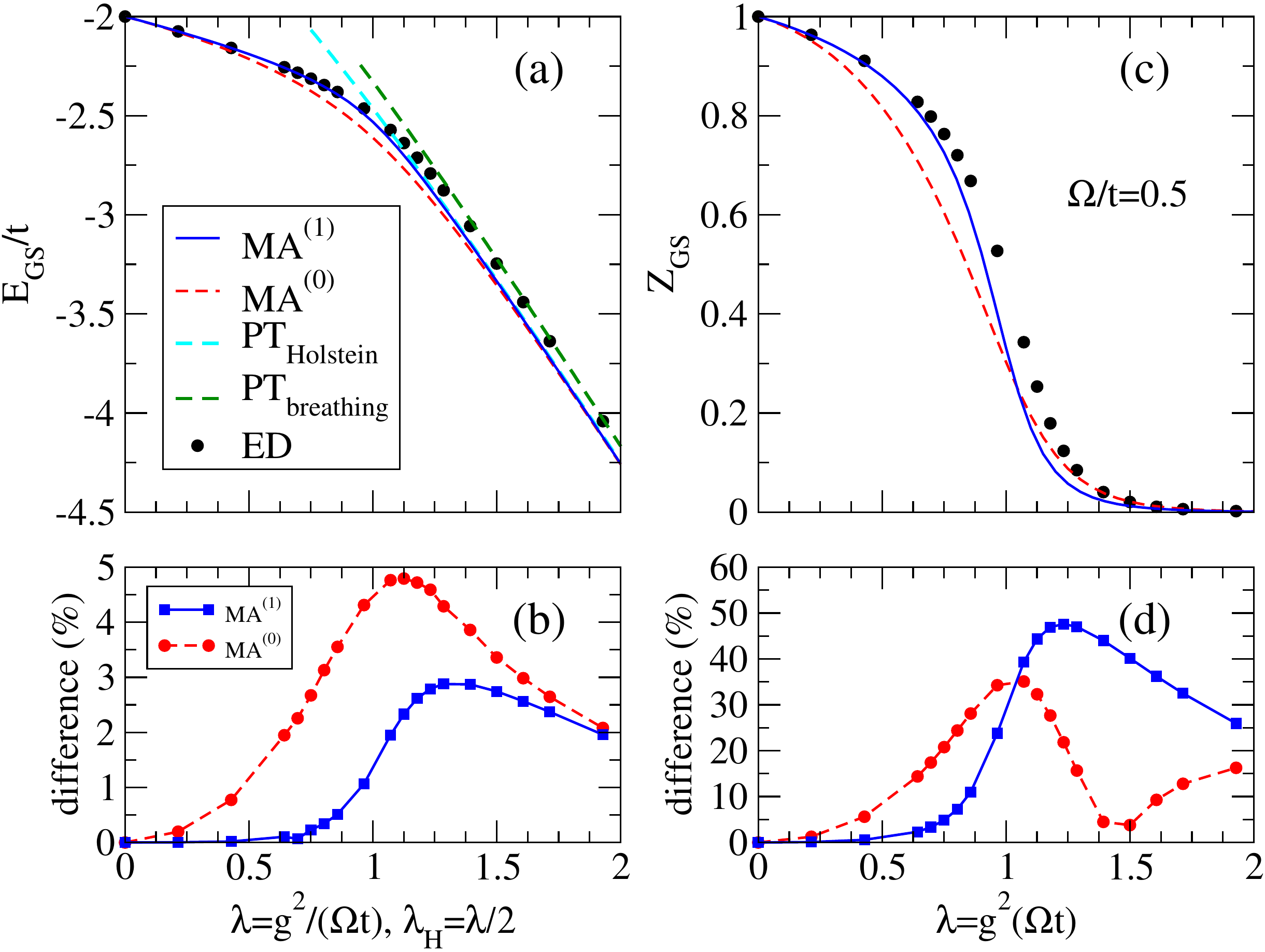}
\caption{(Color online) (a) Ground state energy, (b) Percent
difference from ED ground state energy results, (c) Ground state qp
weight, and (d) Percent difference from ED qp weight results, as a
function of the effective coupling $\lambda$, for $t=1,
\Omega/t=0.5$. The perturbation theory results\cite{lau} for both the
Holstein and breathing-mode models are shown, and the Holstein result
is plotted as a function of $\lambda_H = \lambda /2$. The ED results
are from Ref. \onlinecite{lau}. \label{fig:EZ}}
\end{figure}

In Fig. \ref{fig:EZ} we plot the ground-state energy and qp weight as
a function of the electron-phonon coupling strength, for a phonon
energy $\Omega/t=0.5$.
The MA$^{(0)}$ (red dashed line) and MA$^{(1)}$ (solid blue line)
results both show good agreement with the ED results (black
circles). The ground state energies calculated with MA$^{(0)}$ are
within 5\% error of the ED results, and the MA$^{(1)}$ results are
better, coming within 3\% error of the exact energies. They are exact
for both $t=0$ 
and $\lambda=0$, as expected, and the crossover from the large to
small polaron is captured to a high degree of accuracy. 
This
accuracy is very encouraging, especially since these
approximations are so trivial to evaluate. It is also worth pointing
out that our work on the Holstein Hamiltonian shows that this accuracy
improves in higher-dimensional 
models, and we believe this to be true here as well. Unfortunately,
lack of detailed numerical results in higher dimensions prevents us 
from confirming this to be the case for models with momentum-dependent
coupling. 

Two more observations are apparent regarding these results: (i) The
energies predicted by MA are \emph{lower} in energy than the ED
results, and (ii) the MA results approach the Lang-Firsov asymptotic
limit very slowly.

The fact that the energies predicted are below the exact result
indicates that the MA approximation is non-variational in the case of
the breathing-mode model. This is somewhat surprising, since for the
Holstein model it has been shown that MA$^{(0)}$ is
variational\cite{barisic,berciu3} and that MA$^{(1)}$ is
quasi-variational (in the latter case the Hamiltonian is modified
slightly and the approximation is no longer truly variational, for
details see Ref. \onlinecite{berciu2}). In any case, the energies
found for the Holstein model using the semi-variational MA$^{(1)}$ and
MA$^{(2)}$ approximations were always \emph{higher} than the exact
numerical results.

The slow asymptotic convergence at large $\lambda$ is due to a
different prefactor of the ${\cal O}(t^2)$ perturbational
correction. Instead of the correct breathing-mode result\cite{lau}  
\begin{multline}
E_B(k) = -\frac{2g^2}{\Omega} -2t\cos(ka)e^{-3g^2/\Omega^2} \\ -
\frac{\Omega t^2}{g^2} 
\left[ \frac{1}{3} + \frac{e^{-2g^2/\Omega^2}}{2}\cos(2ka) \right],
\end{multline}
we have produced the Holstein result from Ref. \onlinecite{goodvin}
with $g^2$ replaced with $\bar{g}^2=2g^2$ [see the two dashed lines
  labeled PT in Fig. \ref{fig:EZ}(a)]: 
\begin{multline}
E_H(k) = -\frac{2g^2}{\Omega} -2t\cos(ka)e^{-2g^2/\Omega^2} \\ -
\frac{\Omega t^2}{g^2} 
\left[ \frac{1}{2} + e^{-2g^2/\Omega^2}\cos(2ka) \right].
\end{multline}

To understand the origin for both of these facts, we observe that at this
level of approximations, almost all
dependence on the el-ph coupling $g_\mb{q}$ is through its
momentum-averaged value  $\bar{g}^2=1/N\sum_q |g_q|^2$. For the
1D breathing-mode model, this average happens to be the same whether
$g_q \propto 
i\sin{qa\over 2}$, as is the case here, or whether $g_q \propto
\cos{qa\over 2}$ which would correspond to a coupling of the electron
to the sum $x_{i-{1\over 2}} +x_{i+{1\over 2}}$ of O-site
displacements. In other words, for this model, these simplest versions
of the MA approximation register that a phonon cloud is formed at a
certain O site, but not whether this leads to a leftwards or
rightwards displacement of that site.  In the strong coupling limit,
one expects 
clouds to be formed only on the two O sites neighboring the Cu site
that hosts the charge, and to point towards the charge. The effective
hopping is related to the overlap  of these clouds when the charge
hops to a neighboring site. If the electron hops from $i$ to $i+1$,
the phonon cloud at $i+{1\over 2}$ changes from $\exp{[-{g\over \Omega}
  b^\dagger_{i+{1\over 2}}]}|0\rangle$ to $\exp{[{g\over \Omega}
  b^\dagger_{i+{1\over 2}}]}|0\rangle$ and the overlap is small.  In
the MA approximations, it is precisely this information about left
vs. right that is lost (equivalently, one can think of losing the
information about relative phases between various contributions to the
phonon cloud) and the overlap is unity -- both cases have a cloud on
the $i+{1\over 2}$ site. This explains the higher polaron mobility,
hence the lower energy and slower
convergence towards the exact asymptotic value for the MA
approximations. 

\begin{figure}[t]
\includegraphics[width=0.70\columnwidth]{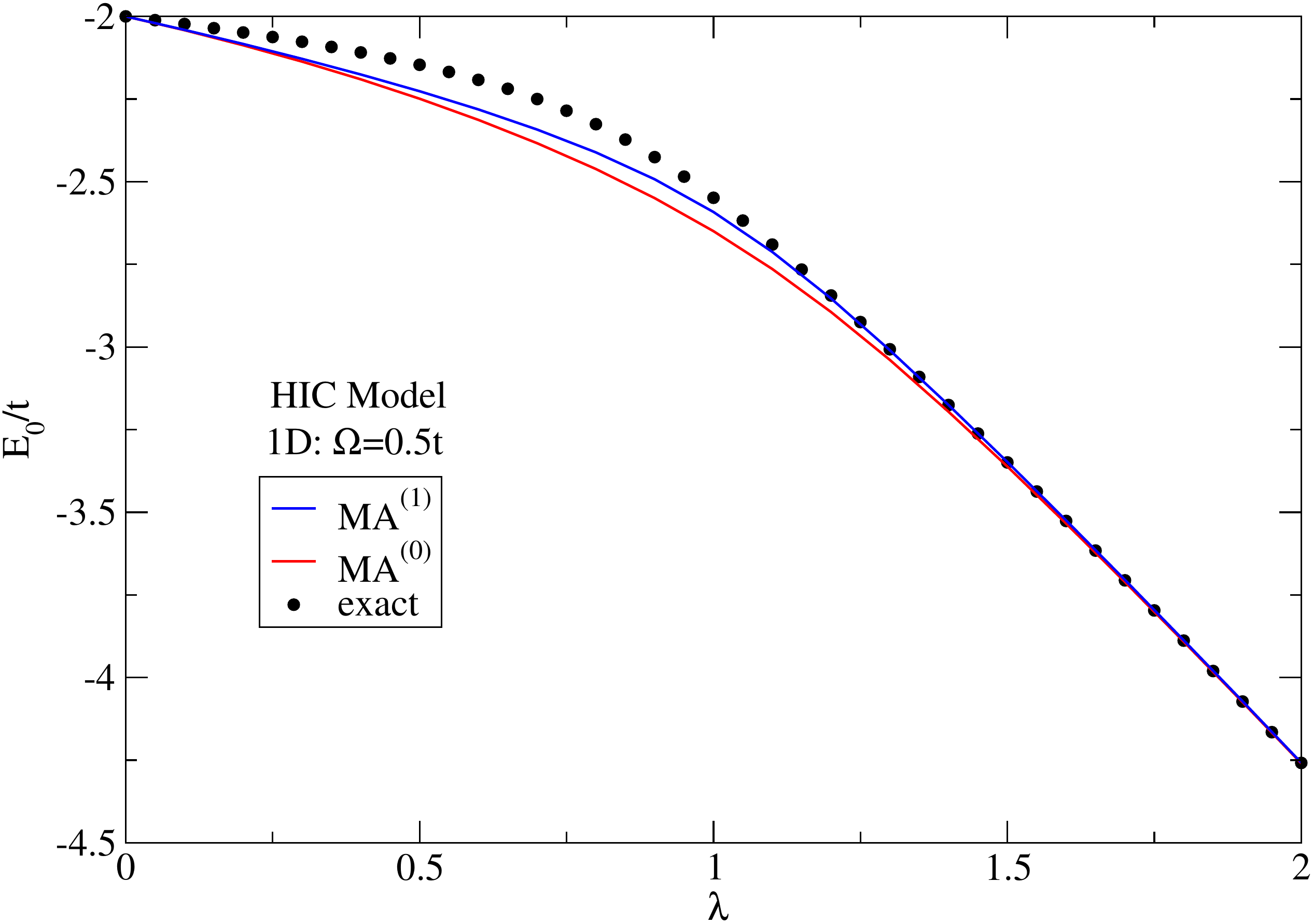}
\caption{MA$^{(0)}$ and
MA$^{(1)}$ predictions of the ground-state energy of the HIC
model\cite{HIC} as compared to its exact solution (black dots). For
this model with infinite-range electron-phonon coupling, the
approximations capture the correct asymptotic behavior.
\label{fig:HIC}}
\end{figure}

For the Holstein model, this problem does not appear because the el-ph
coupling is local. We argue that this problem should also become less
serious if the el-ph coupling is longer range, because in that case
the relative sign of various displacements is not completely lost in
the $\bar{g}^2=1/N\sum_q |g_q|^2$ average.  Unfortunately, lack of
numerical data for lattice models with longer-range el-ph coupling
makes it difficult to support this statement. The only comparison we
can offer is with the exact solution of a somewhat pathological,
infinite-range coupling model called the HIC model of
Ref. \onlinecite{HIC}. As shown in Fig. \ref{fig:HIC}, for this
model both MA$^{(0)}$ and MA$^{(1)}$ approximations
give the correct asymptotic behavior -- and  this necessitates a correct
description of the infinite-range phonon clouds that appear in this
model.

To summarize,  the MA approximations given
by Eqs. (\ref{eq:MA0}) and (\ref{eq:MA1}) are very 
easy-to-use, and rather accurate for models with
momentum-dependent electron-phonon coupling. Given its low dimension
and short-range (but 
not local) el-ph interaction,  one would expect the  1D
breathing-mode model to be amongst the worst examples for the accuracy
of this approximation. However, even here we obtain very decent
agreement. As argued, we expect this to improve for higher dimensions
and longer-range interactions. The only regime where accuracy is
certain to worsen is when $\Omega/t$ becomes very small  (this is a
general problem of  MA-like approximations, discussed at length in
Ref. \onlinecite{goodvin}). Therefore, we believe that these
approximations are  useful for quick guides to relevant energies and other
quantities of interest for problems of this type, obtained 
with minimal analytical and computational
effort, yet still reasonably accurate.

\begin{figure}[t]
\includegraphics[width=0.90\columnwidth]{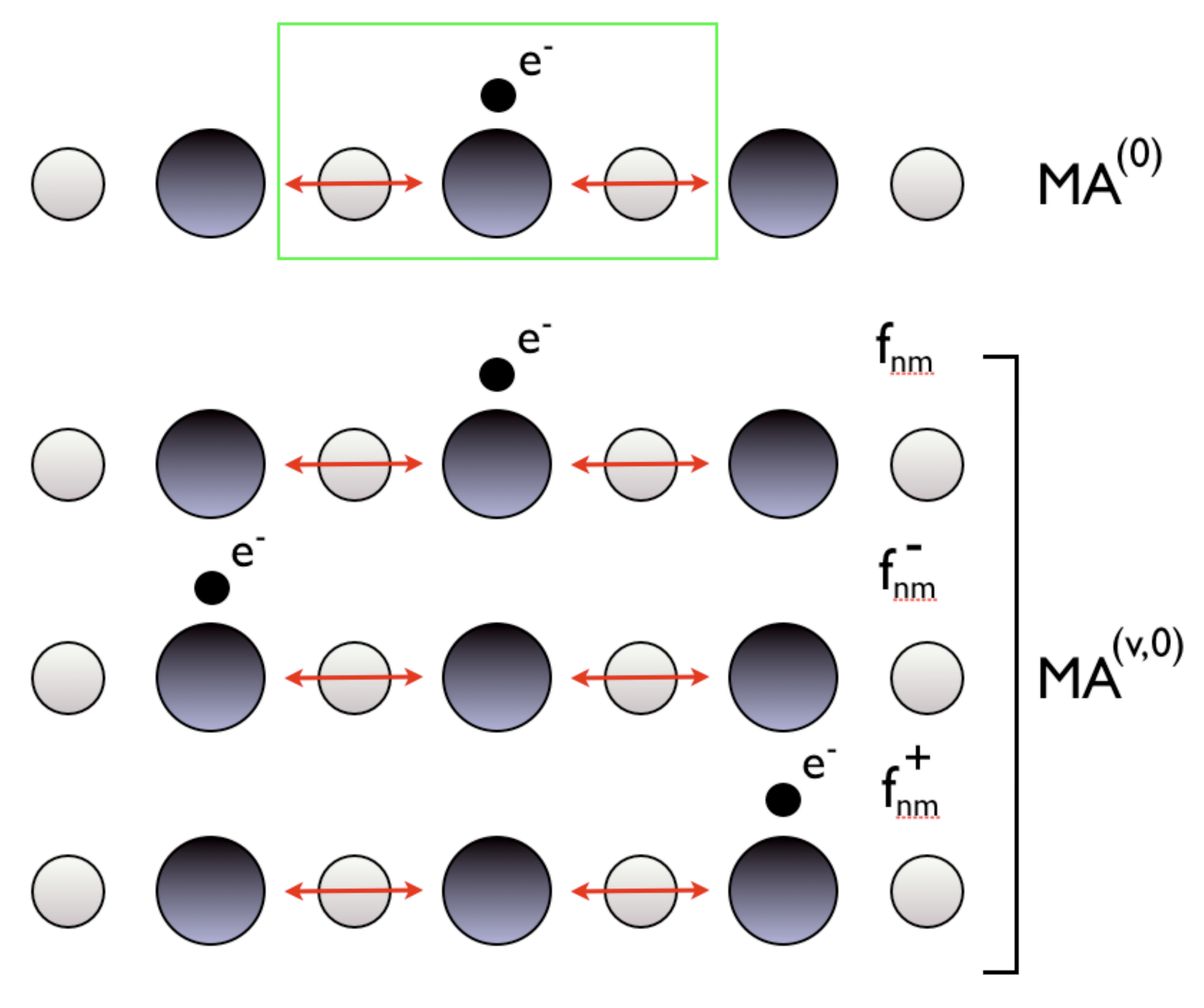}
\caption{The generalized Green's functions that appear in 
  MA$^{(0)}$ are sketched
in the top picture. They have phonon clouds on two neighboring sites,
with the electron on the central site. For the variational MA
approximation, denoted 
MA$^{(v,0)}$,  we find again such Green's functions, but also those
with the electron to the immediate left or right of the phonon
clouds.  We name them
$f_{nm}$ and $f^{\mp}_{nm}$, respectively.
\label{fig:pic}
}
\end{figure}

\section{The variational Momentum Average
  Approximation \label{sec:MAv}}

In this section, we attempt to remedy the problems pointed out in the
previous section and by so doing, to obtain an improved MA
approximation for the 1D breathing-mode Hamiltonian. The solution,
which is presented in this section, can then be used as a template for any
other $g_{\mb q}$ model. The main idea is to try to formulate an MA
approximation which is 
variational in nature, as the original MA is for the Holstein
Hamiltonian.\cite{berciu2,barisic}

\subsection{The MA$^{(v,0)}$ Approximation \label{sec:MA0v}}

Given the good agreement found in
the previous section, 
it is reasonable to use the 
MA$^{(0)}$ solution as a guidance for what states are most relevant to
include. Remember that this solution involved only  the fully
momentum-average quantities ${\cal F}_n ({\mb k}, \omega) = {1\over
  N^n}\sum_{\mb{q}_1, \dots,\mb{q}_n}^{} f_n(\mb{k}, \mb{q}_1,
\dots,\mb{q}_n, \omega)$. Using the definition of Eqs. (\ref{eq:fF})
and (\ref{eq:FF}) and performing the sums over $q_1,\dots q_n$ for our
specific model with $g_q$ given by Eq. (\ref{eq:gq}), we find
immediately that ${\cal F}_n \propto \sum_{i}^{} e^{ikR_i}\langle 0| c_{\mb k}
\hat{G}(\omega) c_i^\dagger (b_{i-{1\over2}}^\dagger - b_{i+{1\over
    2}}^\dagger)^2|0\rangle$. In other words, only states where there
are phonons only on two neighboring sites contribute to it. 

We use this as the criterion for our variational MA
approximation. More specifically,  when we use the Dyson equation to
generate equations of motion, we only keep the terms which have
phonons only on two neighboring sites, and discard any other
contribution. Of course, these states have to have a total momentum
$k$, or else the matrix element is zero. We find that 3 sets of
generalized Green's functions, sketched in Fig. \ref{fig:pic}, 
  appear when we use this
criterion, namely:
\begin{equation} \label{eq:fnmdef}
f_{n,m}=\frac{1}{\sqrt{N}G(k,\omega)}\sum_i e^{ikR_i} \langle0| c_{k}
\hat{G}(\omega)c_i^{\dagger} b_{i-\frac{1}{2}}^{\dagger m}
b_{i+\frac{1}{2}}^{\dagger n-m}|0 \rangle
\end{equation}
and
\begin{equation} \label{eq:fnmpmdef}
f^{\pm}_{n,m}=\frac{1}{\sqrt{N}G(k,\omega)}\sum_i e^{ikR_i} \langle0|
c_{k} \hat{G}(\omega)c_{i\pm1}^{\dagger} b_{i-\frac{1}{2}}^{\dagger m}
b_{i+\frac{1}{2}}^{\dagger n-m}|0 \rangle.
\end{equation}
 Note: as before, we again we
do not write explicitly the $k$ and $\omega$ dependence.

It is straightforward to show that in this notation we can write
Eq. (\ref{eq:G'}) as
\begin{equation}
\label{eq:Gandf} G(k,\omega) = G_0(k,\omega)[ 1+g(f_{1,0} - f_{1,1})G(k,\omega)],
\end{equation}
and therefore
\begin{equation} \label{eq:sigmaMA0v}
\Sigma_{MA^{(v,0)}}(k,\omega) = g(f_{1,0} - f_{1,1}).
\end{equation}
The equations of motion consistent with the variational restriction
are straightforward to find:
\begin{widetext}
\begin{multline}
 \label{eq:fn0}
 f_{n,0} = e_0\left[ nf_{n-1,0} + f_{n+1,0} - f_{n+1,1} \right] - e_2e^{-ika}f^+_{n+1,n} 
 \\ -e_1\left[ne^{-ika}f_{n-1,n-1} - e^{-ika}f_{n+1,n} + e^{-ika}f_{n+1,n+1} - f^-_{n+1,1} \right]     
 \end{multline}
For any $1\le m\le n-1$, 
 \begin{multline}
 f_{n,m} = e_0\left[ (n-m)f_{n-1,m} - mf_{n-1,m-1}  + f_{n+1,m} - f_{n+1,m+1} \right] \\ -e_1 \left[ (n-m)f^+_{n-1,m} - mf^-_{n-1,m-1} + f^+_{n+1,m} - f^-_{n+1,m+1}  \right],     
 \end{multline}
and
 \begin{multline}
 f_{n,n} = -e_0\left[ nf_{n-1,n-1} + f_{n+1,n+1} - f_{n+1,n}  \right] 
 + e_2e^{ika}f^-_{n+1,1} \\ +e_1\left[ne^{ika}f_{n-1,0} - e^{ika}f_{n+1,1} + e^{ika}f_{n+1,0} - f^+_{n+1,n} \right]. 
 \end{multline}
Similarly, for any $1\le m\le n-1$,
 \begin{multline}
 f^{\pm}_{n,m} = e_1\left[ (n-m)f_{n-1,m} - mf_{n-1,m-1}  + f_{n+1,m} - f_{n+1,m+1} \right] \\ -\left\{ \begin{array}{c} e_0 \\ e_2 \end{array} \right\} \left[(n-m)f^+_{n-1,m} + f^+_{n+1,m}  \right] + \left\{ \begin{array}{c} e_2 \\ e_0 \end{array} \right\} \left[ mf^-_{n-1,m-1} + f^-_{n+1,m+1} \right]      
 \end{multline}
and
   \begin{multline}
  \label{eq:f-n0}
 f^-_{n,0} = e_1\left[ nf_{n-1,0} + f_{n+1,0} - f_{n+1,1} \right] + e_0f^-_{n+1,1} 
 - e_3 e^{-ika}f^+_{n+1,n} \\ -e_2\left[ne^{-ika}f_{n-1,n-1} + e^{-ika}f_{n+1,n+1} - e^{-ika}f_{n+1,n}  \right]
 \end{multline}
   \begin{multline} \label{eq:fnnp}
 f^+_{n,n} = -e_1\left[ nf_{n-1,n-1} + f_{n+1,n+1} - f_{n+1,n}  \right] - e_0f^+_{n+1,n} 
 + e_3 e^{ika}f^-_{n+1,1} \\ +e_2\left[ne^{ika}f_{n-1,0} + e^{ika}f_{n+1,0} - e^{ika}f_{n+1,1}  \right]   
 \end{multline}
 \end{widetext}
 where we use the shorthand notation $e_j = g\bar{g}_j(\omega -
 n\Omega)$, and $\bar{g}_j(\omega)$ are real-space Green's functions, defined as
\begin{equation} \label{eq:gj}
\bar{g}_j(\omega) = \frac{1}{N}\sum_q e^{\pm
  iq(ja)}G_0(q,\omega)=G_0(j,0,\omega).
 \end{equation}
The ``$\pm$'' sign in the exponent of Eq. (\ref{eq:gj}) is irrelevant
because $G_0(q,\omega)$ is even with respect to $q$. The explicit
expressions for these and other momentum averaged functions of the
free electron propagator are given in Appendix \ref{sec:averages}
for a tight-binding dispersion. Note that  $f^-_{n,n} =e^{ika}
f_{n,0}$ and $f^+_{n,0} =e^{-ika}
f_{n,0}$, which is why we do not need to keep them as independent
variables.

Equations (\ref{eq:fn0})-(\ref{eq:fnnp}) can now be cast in the form
$\mb{v}_n = \mr{A}_n \mb{v}_{n-1} + \mr{B}_n \mb{v}_{n+1}$, where
 \begin{equation}
\mb{v}_n = \left( f_{n,0}, \ldots, f_{n,n}, f^-_{n,0}, \ldots,
f^-_{n,n-1}, f^+_{n,1}, \ldots, f^+_{n,n} \right)^T
 \end{equation}
collects all generalized Green's functions with a total of $n$
phonons, and the matrices $\mr{A}_n, \mr{B}_n$ are straightforward to
obtain from the equations above. 
The solution of this set of recursive equations  can
then be written as an infinite continued fraction involving products
of matrices:\footnote{An infinite continued fraction solution
involving matrices was also found in the application of MA to the
Holstein model with multiple phonon modes. For details see
Ref. \onlinecite{covaci}.}
\begin{equation}
\label{eq:vn}
\mb{v}_n = \mr{Q}_n\mb{v}_{n-1} = \cfrac{1}{1-\mr{B}_{n}\mr{Q}_{n+1}}\mr{A}_n \mb{v}_{n-1} .
\end{equation}
Although this is a continued fraction of matrices of increasing size,
it is still numerically trivial to evaluate. The dimensions of
$\mr{A}_n$ and $\mr{B}_n$ are $(3n+1) \times (3n-2)$ and
$(3n+1)\times(3n+4)$, respectively, and we determine the truncation
level of the continued fraction using the same criteria as discussed
in the previous section for the MA$^{(0)}$ approximation. In addition,
the $\mr{A}_n$ and $\mr{B}_n$  matrices are very sparse, which
makes multiplication by them  very efficient. Finally, since $v_0 \equiv 1$ by definition, it is
straightforward to calculate $\Sigma_{MA^{(v,0)}}(k,\omega)$ from
Eq. (\ref{eq:sigmaMA0v}) once $v_1 = (f_{1,0}, f_{1,1}, f^-_{1,0},
f^+_{1,1})^T=\mr{Q}_1$ is evaluated from Eq. (\ref{eq:vn}).
 
\subsection{MA$^{(v,1)}$ approximation \label{sec:MA1v}}
We can also systematically improve the variational MA approximation to
 reproduce the electron+phonon
continuum. It is obvious that this will not be predicted by
MA$^{(v,0)}$, since no phonons are allowed to appear far from the main
polaronic cloud. 

We follow the same approach as before: at the MA$^{(v,1)}$ level we keep all
equations involving free electron propagators with $\omega - \Omega$
exactly. In order to work in the enlarged variational space described
by Eqs. (\ref{eq:fnmdef}) and (\ref{eq:fnmpmdef}) we need to define
the following generalized  Green's functions (these are
analogous to the \emph{partial} momentum averages $\bar{{\cal
    F}}(\mb{q}_1,\mb{k},\omega)$): 
\begin{multline} \label{eq:fbarnm}
\bar{f}_{n,m}(q_1)=\frac{g_{q_1}}{\sqrt{N} G(k,\omega)}\sum_i e^{i(k-q_1)R_i}
\\ \times\langle0| c_{k} \hat{G}(\omega)c_i^{\dagger}
b_{q_1}^{\dagger} b_{i-\frac{1}{2}}^{\dagger m}
b_{i+\frac{1}{2}}^{\dagger (n-1)-m}|0 \rangle
\end{multline}
and
\begin{multline} \label{eq:fbarnmpm}
\bar{f}^{\pm}_{n,m}(q_1)=\frac{g_{q_1}}{\sqrt{N}G(k,\omega)} \sum_i
e^{i(k-q_1)R_i} \\ \times \langle0| c_{k}
\hat{G}(\omega)c_{i\pm1}^{\dagger} b_{q_1}^{\dagger}
b_{i-\frac{1}{2}}^{\dagger m} b_{i+\frac{1}{2}}^{\dagger (n-1)-m}|0
\rangle.
\end{multline}
These Green's functions explicitly contain states with one phonon
delocalized away from the main polaronic cloud, {\em i.e.} precisely the type of
states required 
to reproduce the polaron+phonon continuum. In
this notation Eq. (\ref{eq:f1MA1}) can be written as
\begin{equation}
f_1(q_1) = G_0(k-q_1,\omega - \Omega) \left[ |g_{q_1}|^2 + g
(\bar{f}_{2,0}(q_1) - \bar{f}_{2,1}(q_1) ) \right],
\end{equation}
which is an exact equation involving no approximations, as in the
MA$^{(1)}$ case.

We can immediately use the MA$^{(v,0)}$ result to solve
Eq. (\ref{eq:fnMA1}), but only up to the $n=2$ level: $\mb{v}_2 =
\mr{Q}_2 \mb{v}_1$. To solve the $n=1$ equation exactly we will need
to construct a set of recurrence relations involving
Eqs. (\ref{eq:fbarnm}) and (\ref{eq:fbarnmpm}) using Dyson's
identity. The resulting equations take a form similar to
Eqs. (\ref{eq:fn0})-(\ref{eq:fnnp}) if we define
\begin{multline} \label{eq:dfnm}
\delta f_{n,m}(q_1) = \bar{f}_{n,m}(q_1) - g(1-e^{iq_1})f_{n,m} \\ +
g(1-e^{-iq_1})f_{n,m+1},
\end{multline}
\begin{multline} \label{eq:dfn0}
\delta f_{n,0}(q_1) = \bar{f}_{n,0}(q_1) - g(1-e^{iq_1})f_{n,0} \\ +
g(1-e^{-iq_1})f_{n,1} -ge^{-i(k-q_1)}(1-e^{iq_1})f^-_{n,n-1},
\end{multline}
\begin{multline} \label{eq:dfnn}
\delta f_{n,n-1}(q_1) = \bar{f}_{n,n-1}(q_1) - g(1-e^{iq_1})f_{n,n-1}
\\ + g(1-e^{-iq_1})f_{n,n} + ge^{i(k-q_1)}(1-e^{-iq_1})f^+_{n,1},
\end{multline}
for $n>1$, and
\begin{multline} \label{eq:df10}
\delta f_{1,0}(q_1) = \bar{f}_{1,0}(q_1) - g(1-e^{iq_1})f^{(1)}_{1,0}
+ g(1-e^{-iq_1})f^{(1)}_{1,1} \\
-ge^{-i(k-q_1)}(1-e^{iq_1})f^{-(1)}_{1,0}
+ge^{i(k-q_1)}(1-e^{-iq_1})f^{+(1)}_{1,1}.
\end{multline}
We have again added the superscript $(1)$ to distinguish
$f^{(1)}_{1,0},f^{(1)}_{1,1},f^{-(1)}_{1,0},f^{+(1)}_{1,1}$ from the
MA$^{(v,0)}$ expressions. With these definitions, the recurrence
relations for the $\delta f_{n,m}$ functions have precisely the same
form as Eqs. (\ref{eq:fn0})-(\ref{eq:fnnp}) with $n\to n-1$ and
$\omega \to \omega - \Omega$. As before, we write this set of
equations in the form $\delta \mb{v}_n = \mr{A}'_n \delta \mb{v}_{n-1}
+ \mr{B}'_n \delta \mb{v}_{n+1}$, where
\begin{multline}
\delta \mb{v}_n = \left( \delta f_{n,0}, \ldots, \delta f_{n,n-1},
\right. \\ \left.  \delta f^-_{n,0}, \ldots, \delta f^-_{n,n-2},
\delta f^+_{n,1}, \ldots, \delta f^+_{n,n-1} \right)^T,
\end{multline}
and the solution can again be written in terms of an infinite
continued fraction of matrices:
\begin{equation}
\delta \mb{v}_n= \mr{R}_n \delta \mb{v}_{n-1} =
\cfrac{1}{1-\mr{B}'_n\cfrac{1}{1 - \cdots}\mr{A}'_{n+1}}\mr{A}'_n
\delta \mb{v}_{n-1},
\end{equation}
where $\mr{A}'_n(\omega) = \mr{A}_{n-1}(\omega - \Omega)$ and
$\mr{B}'_n(\omega) = \mr{B}_{n-1}(\omega - \Omega)$.

To finish the calculation we will require explicit expressions for
$f^{(1)}_{1,0},f^{(1)}_{1,1},f^{-(1)}_{1,0},$ and
$f^{+(1)}_{1,1}$. Using the definitions in Eqs. (\ref{eq:fnmdef}),
(\ref{eq:fnmpmdef}), (\ref{eq:fbarnm}), and (\ref{eq:fbarnmpm}) it is
straightforward to show that
\begin{multline} \label{eq:f10}
f^{(1)}_{1,0} = \frac{1}{N}
\sum_{q_1}G_0(k-q_1,\omega-\Omega)\frac{1}{1-e^{iq_1}}\\ \times \left[
\bar{f}_{2,0}(q_1) - \bar{f}_{2,1}(q_1)\right] + e_0 -e_1e^{-ika},
\end{multline}
\begin{multline} \label{eq:f11}
f^{(1)}_{1,1} = -\frac{1}{N}
\sum_{q_1}G_0(k-q_1,\omega-\Omega)\frac{1}{1-e^{-iq_1}}\\ \times
\left[ \bar{f}_{2,0}(q_1) - \bar{f}_{2,1}(q_1)\right] - e_0 +
e_1e^{ika},
\end{multline}
\begin{multline} \label{eq:f10m}
f^{-(1)}_{1,0} = \frac{1}{N}
\sum_{q_1}e^{i(k-q_1)}G_0(k-q_1,\omega-\Omega)\frac{1}{1-e^{iq_1}}\\
\times \left[ \bar{f}_{2,0}(q_1) - \bar{f}_{2,1}(q_1)\right] + e_1
-e_2e^{-ika},
\end{multline}
and
\begin{multline} \label{eq:f11p}
f^{+(1)}_{1,1} = -\frac{1}{N}
\sum_{q_1}e^{-i(k-q_1)}G_0(k-q_1,\omega-\Omega)\frac{1}{1-e^{-iq_1}}\\
\times \left[ \bar{f}_{2,0}(q_1) - \bar{f}_{2,1}(q_1)\right] - e_1 +
e_2e^{ika}.
\end{multline}
To combine all of these results and obtain a system of four equations
in the four unknowns
$f^{(1)}_{1,0},f^{(1)}_{1,1},f^{-(1)}_{1,0},f^{+(1)}_{1,1}$, we need
to rewrite $\bar{f}_{2,0}(q_1) - \bar{f}_{2,1}(q_1)$ in terms of these
four quantities. Using the definitions in
Eqs. (\ref{eq:dfn0})-(\ref{eq:df10}), $\mb{v}_2=\mr{Q}_2\mb{v}_1$, and
$\delta \mb{v}_2 = \mr{R}_1 \delta \mb{v}_1$, we find that
\begin{widetext}
\begin{multline} \label{eq:diff}
\bar{f}_{2,0}(q_1) - \bar{f}_{2,1}(q_1) = G_0^{-1}(k-q,\omega -
\Omega)G_0(k-q,\tilde{\tilde{\omega}})\left\{ (R_{00}-R_{10})\left[
G_0(k-q_1,\omega-\Omega)|g_{q_1}|^2 \right. \right. \\ \left. \left.-
g(1-e^{iq_1})f^{(1)}_{1,0} + g(1-e^{-iq_1})f^{(1)}_{1,1}
-ge^{-i(k-q_1)}(1-e^{iq_1})f^{-(1)}_{1,0} +
ge^{i(k-q_1)}(1-e^{-iq_1})f^{+(1)}_{1,1}) \right] \right. \\
\left. +g(1-e^{iq_1}) \mr{Q}_2\mb{v}_1 |_0
-2g\left[1-\cos(q_1)\right] \mr{Q}_2\mb{v}_1 |_1 
+ g(1-e^{-iq_1})  \mr{Q}_2\mb{v}_1 |_2 \right. \\ \left.
+ ge^{-i(k-q_1)}(1-e^{iq_1}) \mr{Q}_2\mb{v}_1 |_4 +
ge^{i(k-q_1)}(1-e^{-iq_1}) \mr{Q}_2\mb{v}_1 |_5 \right\},
\end{multline}
\end{widetext}
where $\tilde{\tilde{\omega}} = \omega - \Omega - g(R_{00}-R_{10})$. In this notation  $R_{ij}$ are the matrix elements of $\rm{R}_1$ and $\mr{Q}_2\mb{v}_1 |_i$ is the $i^{th}$ element of the product between the matrix $\rm{Q}_2$ and the vector $\mb{v}_1$, \emph{i.e.}, $\mr{Q}_2\mb{v}_1 |_i = Q_{i0}f^{(1)}_{1,0} + Q_{i1}f^{(1)}_{1,1} + Q_{i2}f^{-(1)}_{1,0} + Q_{i3}f^{+(1)}_{1,1}$, where $Q_{ij}$ are the matrix elements of $\mr{Q}_2$. 
Inserting Eq. (\ref{eq:diff}) into
Eqs. (\ref{eq:f10})-(\ref{eq:f11p}), and performing the required sums
over momentum, we obtain the desired four equations in four
unknowns. The explicit expressions for the momentum averages that
appear in these expressions are given in Appendix
\ref{sec:averages}. Once this system has been solved we can easily
compute the self-energy as
\begin{equation}
\Sigma_{MA^{(v,1)}}(k,\omega)= g(f^{(1)}_{1,0} - f^{(1)}_{1,1}).
\end{equation}
Again the MA$^{(v,1)}$ expression is slightly more involved than the
zeroth order approximation because we need to evaluate two continued
fractions, but it is still numerically trivial to evaluate.

\section{Results \label{sec:MAvresults}}
We present results for the 1D breathing-mode
Hamiltonian using the variational MA approximation. Comparisons are
made to ED data\cite{lau} where possible. 

\subsection{Ground State Properties}
\begin{figure}[b]
\includegraphics[width=0.90\columnwidth]{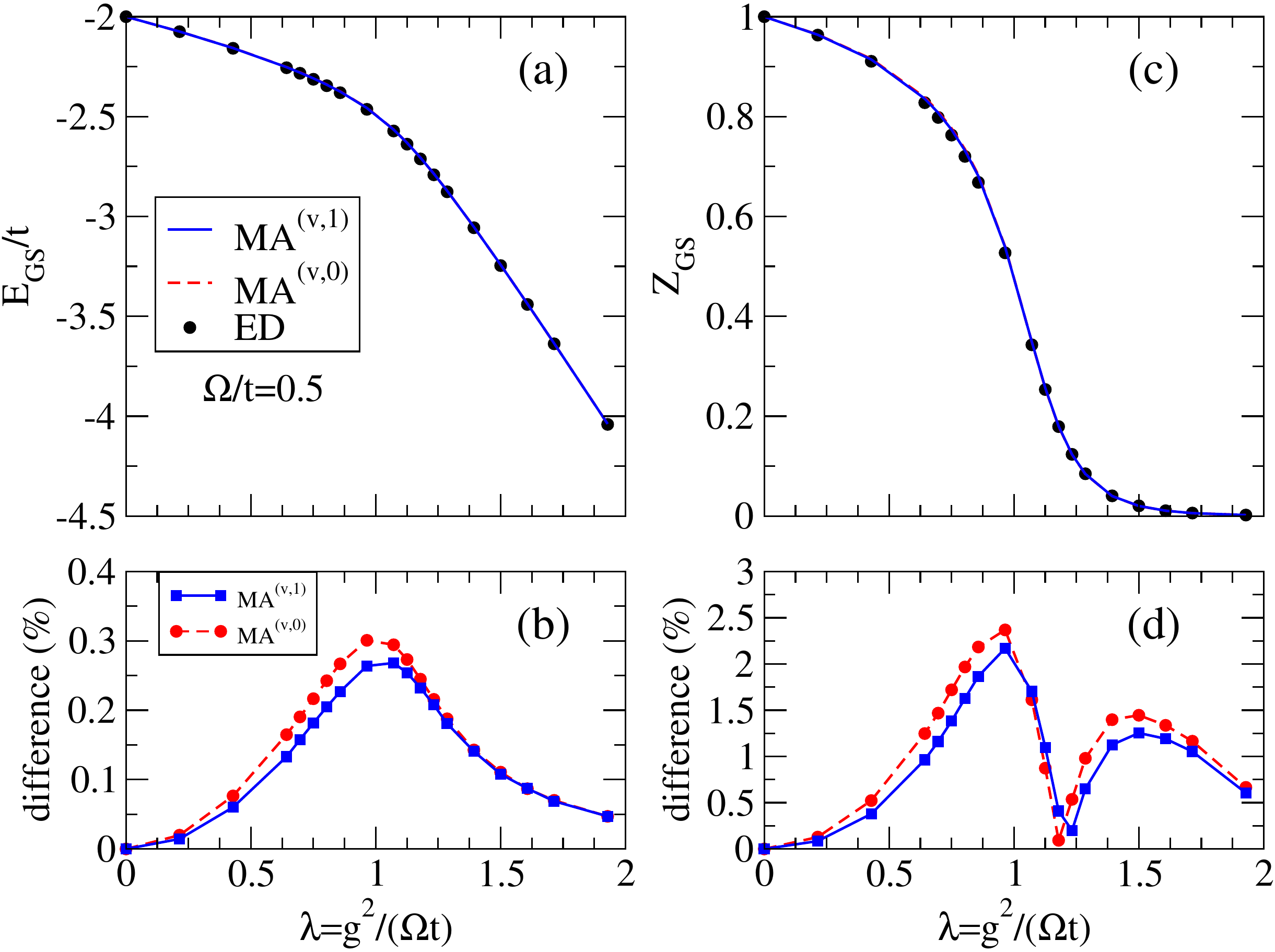}
\caption{(Color online) (a) Ground state energy, (b) Percent
difference from ED ground state energy results, (c) Ground state qp
weight, and (d) Percent difference from ED qp weight results, as a
function of the effective coupling $\lambda$, for $t=1,
\Omega/t=0.5$. The ED results are from Ref. \onlinecite{lau}.
\label{fig:EZv}}
\end{figure}

In Fig. \ref{fig:EZv} we plot both the ground state energy and qp
weight as a function of the electron-phonon coupling strength, using
the variational MA approximation.
The variational MA results show a clear improvement over the MA
results shown in Fig. \ref{fig:EZ}, and the agreement with the
numerical data is excellent. The approximate and exact numerical data
results are indistinguishable when plotted over the full parameter
range of Fig. \ref{fig:EZv}(a). To gain a closer look at the success
of our approximation we plot the relative error between the MA results
and the ED results in Fig. \ref{fig:EZv}(b). Indeed the approximation
is giving extremely close agreement with the numerical results, with
less than 0.3\% relative error for both MA$^{(v,0)}$ and
MA$^{(v,1)}$. The largest errors occur at intermediate couplings, as
expected because the MA approximation is exact in both the zero
coupling and zero bandwidth limits. We also show a comparison of the
quasiparticle weight calculated using variational MA with the ED
result in Fig. \ref{fig:EZv}(c). Again, when plotted over the full
parameter range, the results are nearly indistinguishable. A look at
the percent difference indicates that the relative error is less than
2.5\%. This is a truly remarkable fact considering that the qp weight
contains information on the nature of the eigenstates, something that
is rarely obtained accurately when using approximate methods.

It is clear from Fig. \ref{fig:EZv} that the variational MA
approximation cures all of the shortcomings of the simple MA
generalization discussed in Sec. \ref{sec:MAresults}. A comparison of
the variational MA and ED results shows that the MA energies are
slightly higher than the exact numerical results, as expected from a
variational method, and that the asymptotic behavior predicted from
both the perturbational theory and ED result is reproduced. These
successes were expected because the variational MA approximation was
designed precisely to remedy these problems, but it is still very
encouraging to see that the physical picture described and used to
motivate the variational MA approximation in Sec. \ref{sec:MAv} was
indeed correct.

\begin{figure}
\includegraphics[width=0.80\columnwidth]{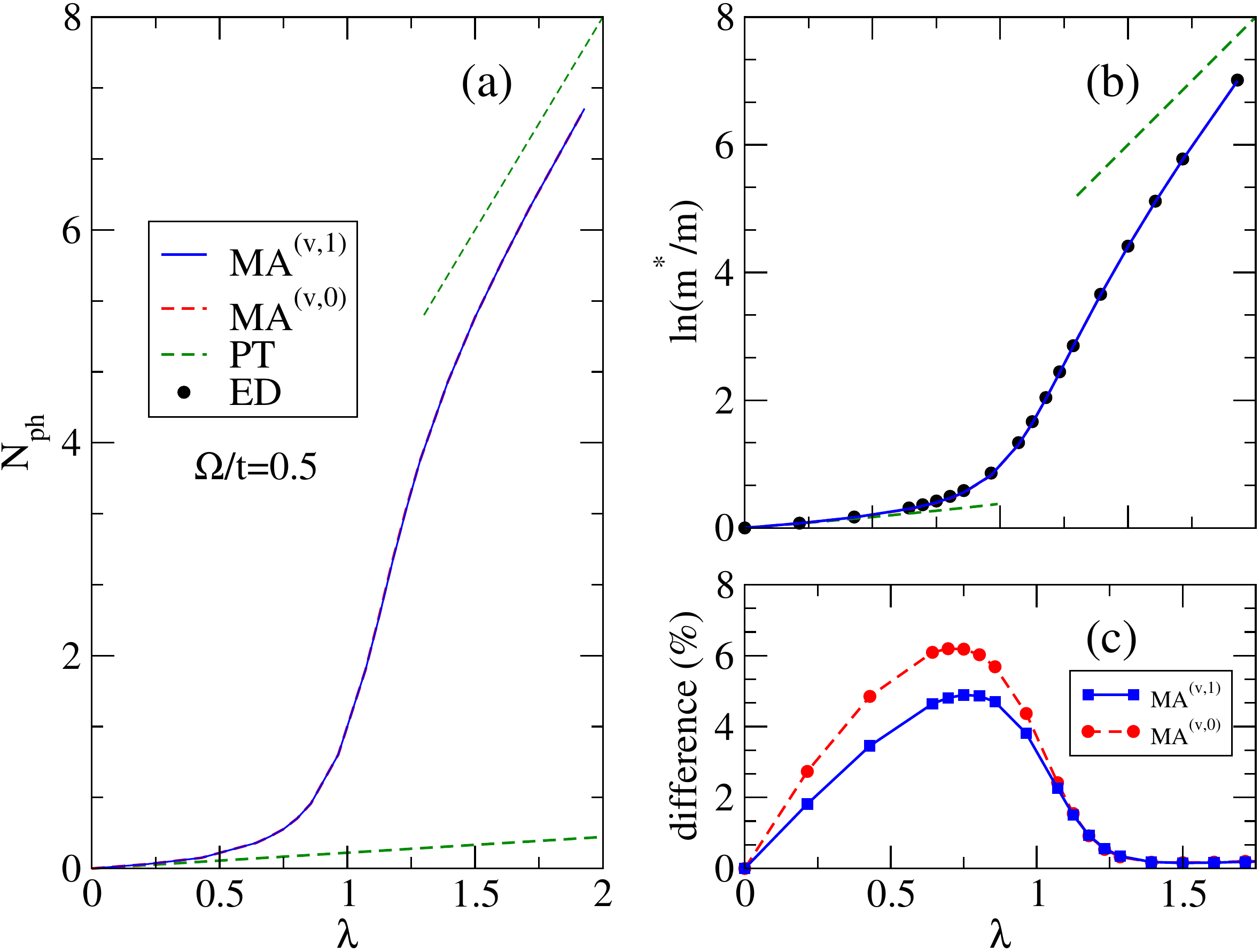}
\caption{(Color online) (a) Average number of phonons and (b)
effective mass, (c) Percent difference from ED effective mass results,
as a function of the effective coupling $\lambda$, for $t=1,
\Omega/t=0.5$. The ED results are from Ref. \onlinecite{lau}.
\label{fig:Nlogm}}
\end{figure}

In Fig. \ref{fig:Nlogm}(a) we plot the average number of phonons in
the cloud, $N_{ph}$, as calculated from the Hellmann-Feynman
theorem.\cite{goodvin} Since there is no numerical data available for
this quantity we compare our findings to the standard
Rayleigh-Schr\"odinger (RS) perturbation theory at small couplings,
and to the strong-coupling perturbation theory for larger
couplings. The approximation reproduces both asymptotic limits to a
high degree of accuracy, as should be expected based on the success of
the ground state energies and qp weights shown in
Fig. \ref{fig:EZv}. Since we are unaware of any explicit expression
for the RS energy for the breathing-mode Hamiltonian in the
literature, we state the result here. By evaluating the standard
expression for the first order correction
\begin{equation}
E_{\mb{k}}^{(1)} = - \frac{1}{N}\sum_{\mb{q}}
\frac{g_{\mb{q}}^2}{\varepsilon_{\mb{k}-\mb{q}} + \Omega -
\varepsilon_{\mb{k}}},
\end{equation}
one can easily show that
\begin{equation}
E_k^{(1)} = -2g^2\bar{g}_0(\Omega - \varepsilon_k) - \frac{g^2}{t}\cos
k \left[ 1 - (\Omega - \varepsilon_k) \bar{g}_0(\Omega -
\varepsilon_k) \right ]
\end{equation}
for the 1D breathing-mode Hamiltonian.

In Fig. \ref{fig:Nlogm} we also plot the effective mass as a function
of the coupling strength.
Again the agreement with the numerical data is excellent, as confirmed
by the relative errors shown in 
Fig. \ref{fig:Nlogm}(c).

\subsection{The Polaron Band}
\begin{figure}[b]
\includegraphics[width=0.90\columnwidth]{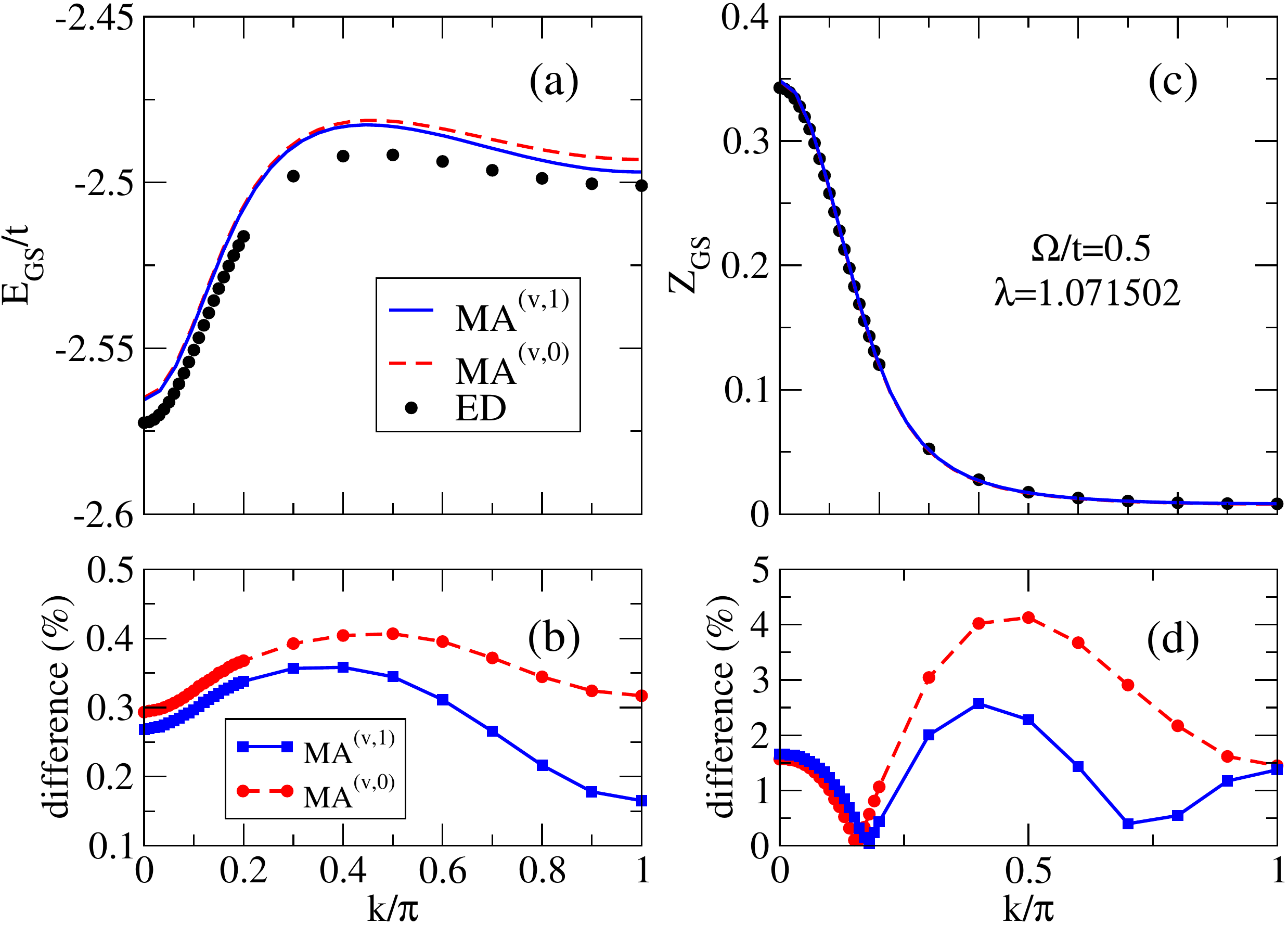}
\caption{(Color online) (a) Polaron dispersion $E_k$, (b) Percent
difference from ED ground state energy results, (c) quasiparticle
weight $Z_k$, and (d) Percent difference from ED qp weight
results. Results are shown for $t=1, \Omega/t=0.5$, and
$\lambda=1.071502$. The ED results are from Ref. \onlinecite{lau}.
\label{fig:EZk}}
\end{figure}

With our analytical expression for the self-energy we can also
calculate momentum-dependent results. In Fig. \ref{fig:EZk}(a) we plot
the lowest energy state for momenta $0 \leq k \leq \pi$ and compare
our results to the available numerical data.
We again find that the variational MA approximation is highly
accurate. Because the polaron dispersion is relatively flat for the
intermediate coupling strength of $\lambda=1.071502$ shown here, the
energy range of the plot is quite narrow and we can clearly discern
the difference between the approximate and numerical results. However,
by looking at the relative error in Fig. \ref{fig:EZk}(b), we
demonstrate that the accuracy of the variational MA result is again
very good, coming well within 0.5\% relative error of the numerical
result, showing that the variational MA approximation is accurate
for all momenta $k$. The reason for the non-monotonic polaron
dispersion has been discussed at length in Ref. \onlinecite{lau}, and is
due to a larger effective 2nd nearest neighbor hopping than the
effective nearest neighbor hopping of the polaron -- this is a direct
consequence of the structure of the polaronic cloud.
In Figs. \ref{fig:EZk}(c) and \ref{fig:EZk}(d) we
plot the qp weight and its relative error as a function of momenta. We
again find good  agreement. As explaned, the agreement is expected to
improve for both smaller and larger $\lambda$ values, where
our approximations become asymptotically exact.

\subsection{Higher Energy Properties}
\begin{figure}[t]
\includegraphics[width=0.90\columnwidth]{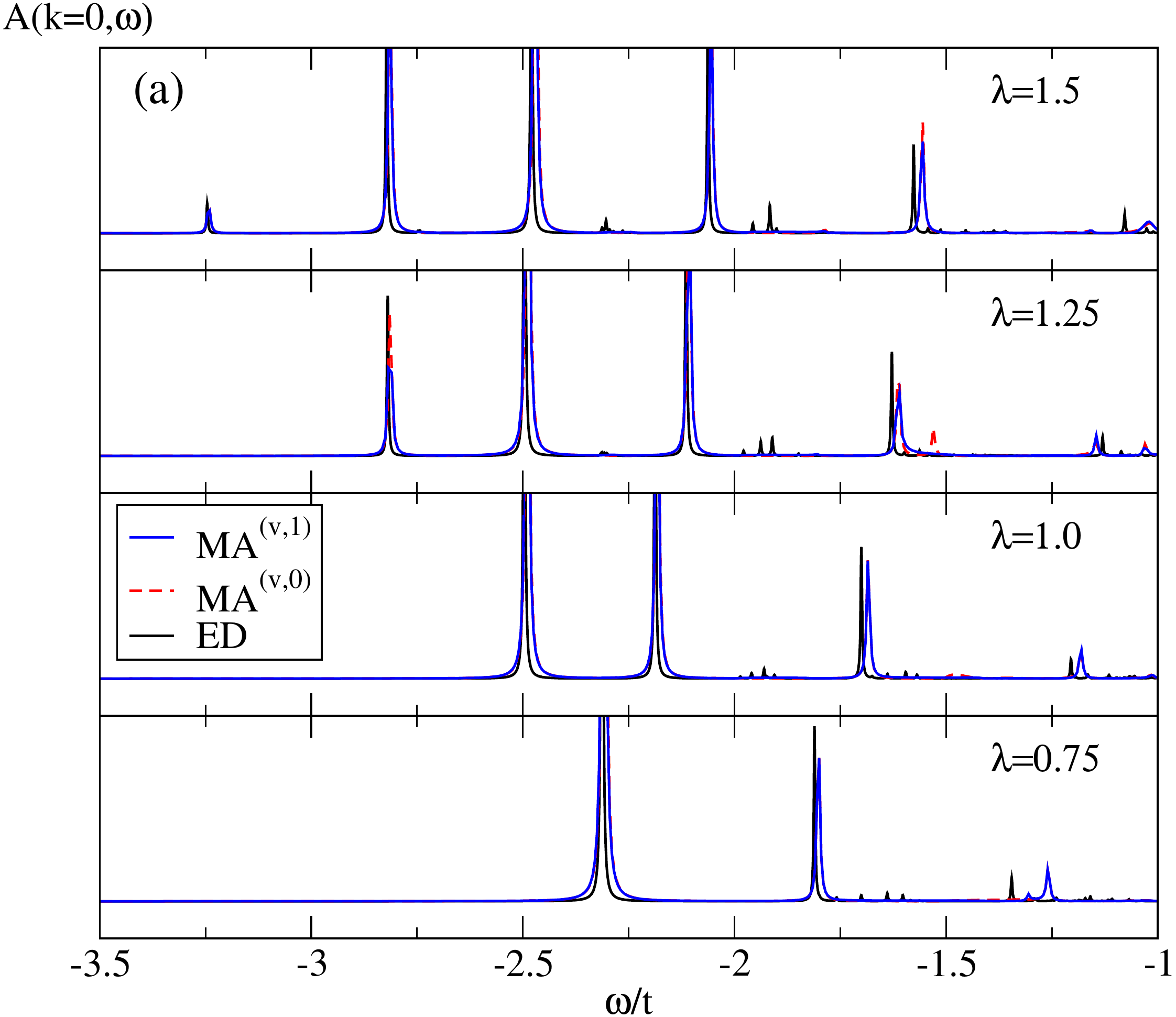}
\includegraphics[width=0.90\columnwidth]{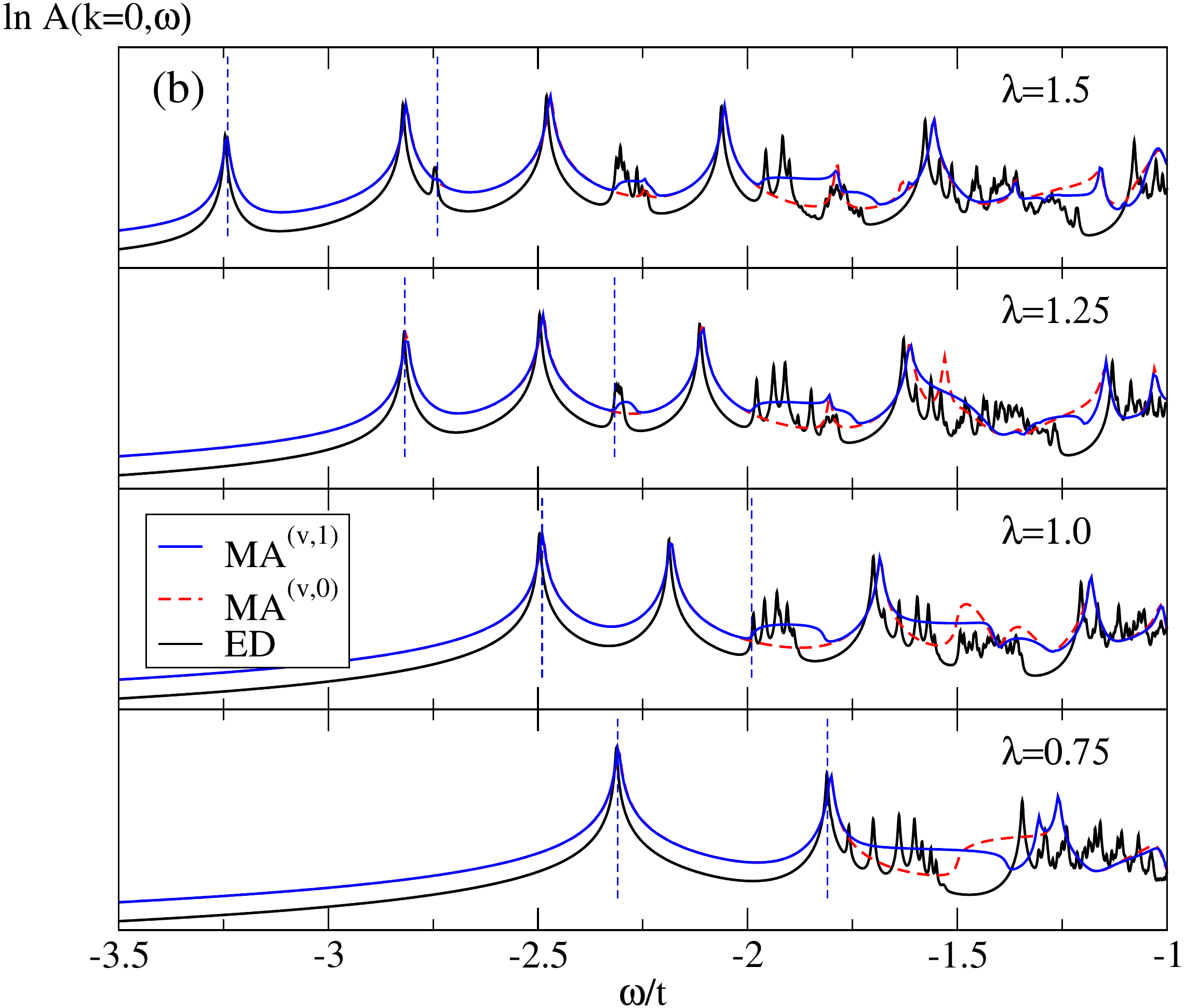}
\caption{(Color online) (a) $A(k=0,\omega)$ vs. $\omega$ and (b) $\ln
A(k=0,\omega)$ for $t=1, \Omega=0.5, \eta=0.004$, and $\lambda =
0.75,1.0,1.25,1.5$; The vertical blue dashed lines denote $E_{gs}$ and
$E_{gs}+\Omega$ using MA$^{(v,1)}$. The ED results are from
Ref. \onlinecite{lau}.
\label{fig:Aw}}
\end{figure}

Lastly, we consider the high energy  properties of the 1D
breathing-mode Hamiltonian using MA$^{(v,0)}$ and MA$^{(v,1)}$. In
Fig. \ref{fig:Aw} we compare our predicted spectral weights to  available
numerical data.
When plotted on a linear axis, the results are essentially indistinguishable,
especially for MA$^{(v,1)}$. To gain a better view we display the same
plots on a logarithmic scale in 
Fig. \ref{fig:Aw}(b). There are a few interesting features to note. At
the MA$^{(0)}$ level (red dashed line) the 
majority of the spectral weight is found to be in the correct
location, however, the electron+phonon continuum is completely absent
for the parameters shown, as expected. At the MA$^{(v,1)}$ level the
continuum is reproduced in the 
expected location, in very good  agreement with the ED prediction.
Furthermore, the finite size effects responsible for the sharp peaks
in the continuum of the ED result are  absent from
the MA$^{(v,1)}$ data. As a guide illustrating the correct location
for the polaron+phonon continuum we have added the vertical blue
dashed lines to denote $E_{gs}$ and $E_{gs}+\Omega$ for the
MA$^{(v,1)}$ ground state energy. The lower edge is located correctly,
however especially for smaller $\lambda$, MA predicts 
a somewhat wider continuum than ED. Given the very limited
availability of numerical
results of this type, we do not know if this discrepancy is due to
truncation approximations in the ED solution, or is due to
inaccuracies of the MA approximations. 

\begin{figure}[htb]
\includegraphics[width=0.90\columnwidth]{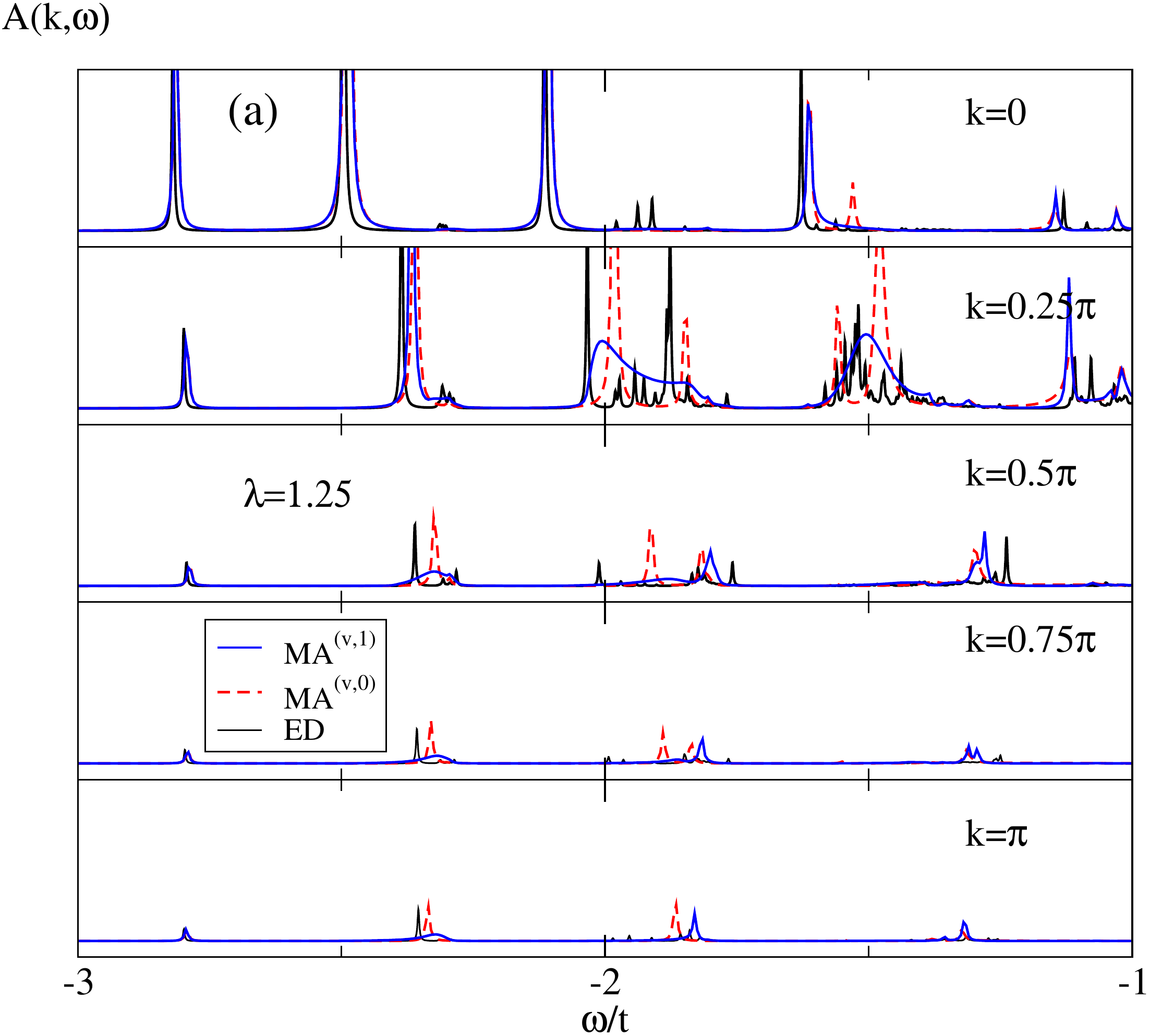}
\includegraphics[width=0.90\columnwidth]{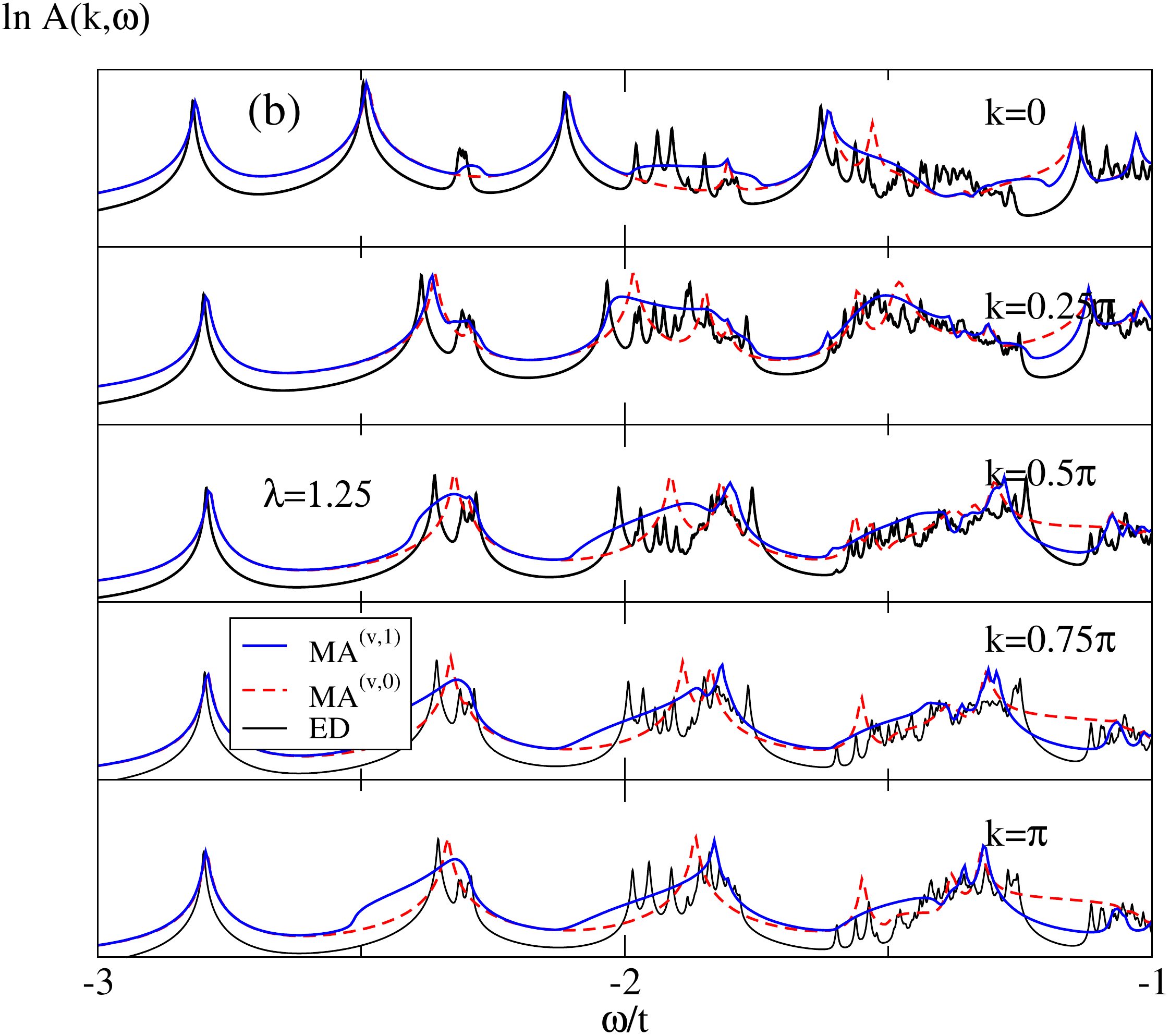}
\caption{(Color online) (a) $A(k,\omega)$ and (b) $\ln A(k,\omega)$
vs. $\omega$ for $k/\pi=0, 0.25, 0.5, 0.75,1$, $t=1, \Omega=0.5,
\lambda = 1.25, \eta=0.004$; The ED results are from
Ref. \onlinecite{lau}.
\label{fig:Akw}}
\end{figure}

In Fig. \ref{fig:Akw} we plot the spectral weight for a fixed coupling
strength and vary the momentum $k$.
Again on the linear scale the results are very encouraging. The MA
result, particularly MA$^{(v,1)}$ predicts the correct location for
the spectral weight over the entire energy range. However,
a logarithmic scale plot does reveal a notable shortcoming of the
approximation. For $k=0$ the agreement with the numerical data is
excellent and the continuum is located at the expected
$E_{gs}+\Omega$. For $k=0.25\pi$ and $k=0.5\pi$ the agreement is still
good, but as we begin to approach the band edge we see a significant
deviation of the MA result from the ED result, with the MA continuum
coming to much too low energies. This feature has extremely
little spectral 
weight (see Fig. \ref{fig:Akw}(a)) and is extremely unlikely to be
detectable in any experimental realization, however it is in stark
disagreement with the expected 
result confirmed by the ED data. We do not currently understand this failure at
higher $k$ values.

\begin{figure}[t]
\includegraphics[width=0.90\columnwidth]{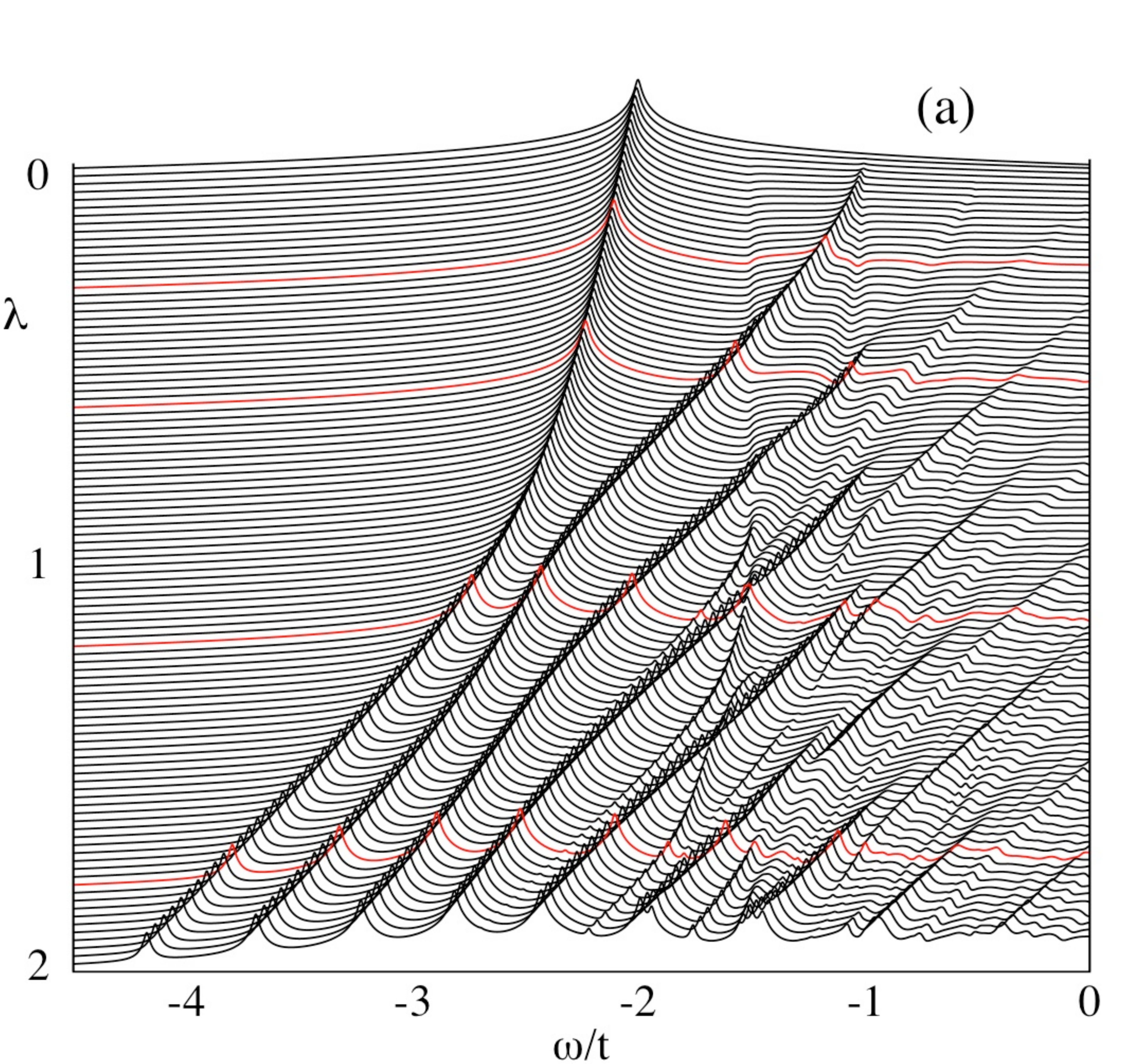}
\includegraphics[width=0.90\columnwidth]{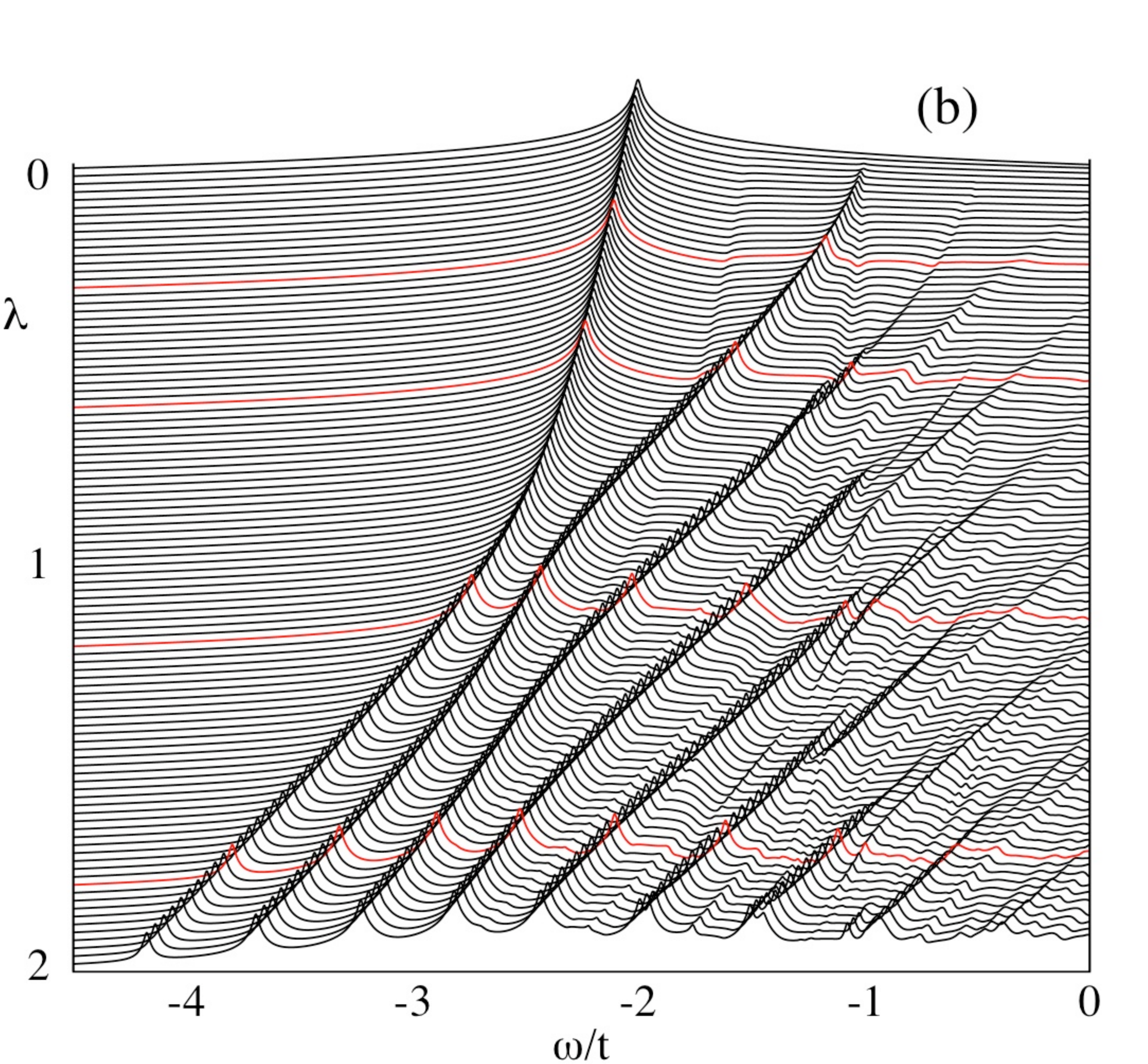}
\caption{(Color online) Spectral weight $A(0,\omega)$ vs. $\omega$ in
1D using (a) MA$^{(v,0)}$ and (b) MA$^{(v,1)}$. The results are shown
for $t=1, \Omega=0.5,\eta=0.01$, and $\lambda$ varying from 0 to
2. Curves corresponding to $\lambda=0.3, 0.6, 1.2$ and $1.8$ are
highlighted in red. \label{fig:contour}}
\end{figure}

Finally, in Fig. \ref{fig:contour} we show more detailed plots of the
evolution of the spectral weight for $k=0$ from zero coupling to large
coupling, for a full range of energies. The two panels compare the
MA$^{(v,0)}$ (a) and MA$^{(v,1)}$ (b) solutions. The one obvious
difference is the location of the polaron+phonon continuum, which is
incorectly located around $-2t$ for MA$^{(v,0)}$, whereas it always
starts at $\Omega$ above $E_{gs}$ for MA$^{(v,1)}$. Aside from this
feature with rather little weight, all the features which contribute
significantly to the spectral weight are in good agreement.

Plots of this nature would require extensive numerical computational
time if one tried to generate them using ``exact'' computational
methods. For example, achieving convergence for the ED results 
at larger $\lambda$ required inclusion of billions of states, in some
cases, in the truncated Hilbert space. Diagonalizing such large (even
though sparse) matrices is not a trivial task. By comparison, because
we have an analytical expression for the 
self-energy, these detailed MA plots covering many energies and couplings
take just seconds, at most  minutes, to generate. In fact, MA is much
faster even than SCBA, which for this model requires many numerical integrals
over the Brillouin zone (and is of questionable use for medium and
large couplings).

\section{Summary and Conclusions}
In summary, we have presented a way to generalize the momentum average
approximation to more general electron-phonon coupling models with
momentum-dependent couplings. The approximation still sums all of the
self-energy diagrams, albeit with each diagram approximated in such a
way that the full sum can be performed, and it is exact in both the
zero coupling and zero bandwidth limits. As in the application of MA
to the Holstein model, the approximation is analytical and
easy-to-use, and gives highly accurate results over the entire
parameter space. In this paper we have actually presented two
different generalizations of MA, a straightforward generalization that
can easily be applied to any model with a momentum-dependent coupling
with minimal effort, and a more specific (and extremely accurate)
generalization that takes the details of a given model into account
and enlarges the variational subspace accordingly. We have used the 1D
breathing-mode Hamiltonian as an example to gauge the accuracy of both
of these MA generalizations. While the straightforward extension of MA
gave fairly accurate results, it was non-variational and it approached
the zero-bandwidth asymptotic limit very slowly. The variational MA
approximation remedied both of these problems and produced extremely
accurate results, coming well within 0.3\% error of the available
numerical results for the ground state energies of the breathing-mode
polaron. We also showed that MA can be systematically improved in both
cases, leading to higher accuracy and the correct location for the
electron+phonon continuum.

The successful generalization of MA to this much broader class of
models in very encouraging. Numerical studies of models as complicated
as even the 1D breathing-mode Hamiltonian are very intensive, and the
MA approximation provides a quick and easy-to-use way to gain an
understanding of these more realistic models without having to do a
detailed numerical analysis from the start. We hope that this tool
will be extremely useful for probing more realistic and experimental
realizations of interesting physical systems. The range of
applicability of the MA approximation has been growing steadily. As
mentioned previously, it has been successfully applied to systems with
multiple phonon modes,\cite{covaci} and multiple free-electron
bands.\cite{covaci2} Other obvious and very useful
generalizations of this approximation would be to apply it to finite
particle densities and/or finite temperatures, and possibly to an even
broader class of Hamiltonians involving electron-electron
interactions.

\acknowledgements We thank George A. Sawatzky and Bayo Lau for many
useful discussions and for sharing their numerical results.  This work was
supported by NSERC, CFI, CIfAR Nanoelectronics, Killam Trusts (G.G.),
and the Alfred P. Sloan Foundation.  \appendix

\section{\label{sec:averages}Momentum Averages}
In this section we derive the various momentum averages required to
evaluate the self-energy expressions derived in this work. We require
the following weighted momentum averages of the free electron
propagator for a 1D tight-binding dispersion:
\begin{equation}
\bar{g}_j (\omega)\equiv \frac{1}{N}\sum_q e^{\pm ik(ja)}G_0(q,\omega).
 \end{equation}
 It is straightforward to show that
 \begin{equation}
 \bar{g}_0(\omega) = \frac{\textrm{sgn}(\omega + i\eta)}{\sqrt{(\omega
 + i\eta)^2 - 4t^2}},
 \end{equation}
 \begin{equation}
 \bar{g}_1(\omega) = \frac{1}{2t} \left[ 1-(\omega +
 i\eta)\bar{g}_0(\omega)\right],
 \end{equation}
 \begin{equation}
  \bar{g}_2(\omega) = -\bar{g}_0(\omega) -\frac{\omega + i\eta}{t}
  \bar{g}_1(\omega),
 \end{equation}
 and
  \begin{equation}
  \bar{g}_3(\omega) = \frac{1}{t} -\left[ 3 - \frac{(\omega +
  i\eta)^2}{t^2} \right] \bar{g}_1(\omega). \end{equation}
  
 For the MA$^{(0)}$ and MA$^{(1)}$ self-energy expressions we also
 require $\bar{\bar{g}}_0(\mb{k},\omega)$, as defined in
 Eq. (\ref{eq:g0barbar}). For the 1D breathing-mode model one can show
 that
  \begin{equation}
  \bar{\bar{g}}_0(k,\omega) = 2g^2 \bar{g}_0(\omega) - \frac{g^2 \cos
  k}{t} \left[ 1 - (\omega + i\eta)\bar{g}_0(\omega) \right].
  \end{equation}
  
 In the MA$^{(v,1)}$ calculation we require the momentum average of
 Eq. (\ref{eq:diff}). For the 1D breathing-mode Hamiltonian
 (\emph{i.e.} $g_q=-2ig\sin(q/2)$ and $\varepsilon_k = -2t\cos(k)$) we
 require the following momentum averages:
  \begin{equation}
 \frac{1}{N}\sum_q G_0(k-q,\omega)\frac{e^{\pm ni(k-q)}(1-e^{\mp
 iq})}{1-e^{\pm iq}} =-e^{\mp ik}\bar{g}_{n+1}(\omega),
\end{equation}
  \begin{multline}
 \frac{1}{N}\sum_q G_0(k-q,\tilde{\omega})G_0(k-q,\omega)\frac{e^{\pm
 n i(k-q)}|g_q|^2}{1-e^{\pm iq}} \\ = \frac{g^2}{\omega -
 \tilde{\omega}} \left\{ \bar{g}_n(\tilde{\omega}) - \bar{g}_n(\omega)
 - e^{\mp ik} \left[ \bar{g}_{n+1}(\tilde{\omega}) -
 \bar{g}_{n+1}(\omega) \right] \right\},
 \end{multline}
 and
  \begin{multline}
 \frac{1}{N}\sum_q G_0(k-q,\omega)\frac{e^{\pm ni(k-q)}(1-\cos
 q)}{1-e^{\pm iq}} \\ =\frac{1}{2}\left[ \bar{g}_n(\omega) - e^{\mp
 ik}\bar{g}_{n+1}(\omega) \right].
 \end{multline}


\end{document}